# Model human heart or brain signals

Çağlar Tuncay
Department of Physics, Middle East Technical University, 06531 Ankara, Turkey
caglart@metu.edu.tr

**Abstract:** A new model is suggested and used to mimic various spatial or temporal designs in biological or non biological formations where the focus is on the normal or irregular electrical signals coming from human heart (ECG) or brain (EEG). The electrical activities in several muscles (EMG) or neurons or other organs of human or various animals, such as lobster pyloric neuron, guinea pig inferior olivary neuron, sepia giant axon and mouse neocortical pyramidal neuron and some spatial formations are also considered (in Appendix).

In the biological applications, several elements (cells or tissues) in an organ are taken as various entries in a representative lattice (mesh) where the entries are connected to each other in terms of some molecular diffusions or electrical potential differences. The biological elements evolve in time (with the given tissue or organ) in terms of the mentioned connections (interactions) besides some individual feedings. The anatomical diversity of the species (or organs) is handled in terms of various combinations of the assumptions and parameters for the initial conditions, the connections and the feeding terms and so on. A small number of (iterative) secular equations (coupled map) are solved for the results with few parameters for each case. The same equations may be used for simulation if random parameters are involved. The model, with simple mathematics and easy software, may be followed besides or instead of the known theoretical approaches.

The basic aim of the present contribution is to mimic various empirical data for some electrical activities of human heart or brain (or various animals). The mentioned empirical data are available in various experimental literatures (which are cited or the open access data are borrowed) and the model results may be considered as in good agreement with them.

## 1. Introduction:

It is known that the shapes (patterns) of various biological units and the electrical signals coming out of them are important. (Reference [1] is a limited selection about the related issues.) It is clear that various irregularities within the mentioned designs may be carrying signatures of several illnesses and understanding the underlying mechanisms for the regular or irregular spatial or temporal formations is of vital importance [2, 3, 5-7, 9]. Here biological and temporal formations for model human heart or brain are considered in the main terms and some spatial but non biological or further temporal and biological formations are treated in Appendix.

For the biological and temporal formations: A group of (biological) units (cells or tissues) are represented by the entries of a lattice where the (mutual) interactions are taken into account in terms of the connections (diffusions of nutrients, morphogens, ingredients and so on) between the entries of the lattice and each unit may also be effected individually at a time by the same or similar agents such as the activators or inhibitors (catalyzes) or morphogenesis and so on. Several parameters of the given group of biological units (biological system) are computed in terms of iterations, after assuming several processes for the time evolution of the patterns. Secondly, the bio-electrical signals such as the electrocardiograms (ECG), electroencephalograms (EEG), electrocorticograms (ECoG), electromyograms (EMG) etc., may be handled within the same approach, where the mentioned electrical properties may be product of the same (or similar) processes which govern the spatial formations.

The results of the model may be helpful for investigating some possible clues for the causal mechanisms shaping the regular or irregular patterns and bio-signals. The model is simple and it involves no differential equation; the related mathematics is algebra and the given algorithm for the computations (or simulations) is easy. The model is presented in the following section and the applications with the results are given in the next one or Appendix. The last section is devoted for discussion and conclusion. Please note that, discussing the related literature or making comments about the biological reasons an so on are kept beyond the scope of the present contribution where the model human heart or brain signals (regular or irregular) are focused on.

**2. Model:** It is known that the biological elements (cells, tissues) are not identical as the electrons, atoms, etc. but each of them is unique; yet, each may be similar to the neighbor ones. Thus, many (if not all) biological aspects of each unit (cell, tissue in an organ, etc.) may be approximated as an average of these of the neighbors within various ranges.

The biological individuals are considered as the entries (I,J) of a square (NxN) lattice C(I,J;T) (with I≤N, J≤N) at a time T. In the iterative interaction tours (time, T) the entry (I,J) interact with various neighbor (nn) entries (K,L) and the parameters for C(I,J;T) average in the mean time. (Reference [4] is a limited selection of the related literature about the model.) Furthermore, some feeding terms which represent the growth (or decay) of the units may be changing in time T and it may be taken as variable; F(I,J;T). Hence, the biological mood (state) of the unit at (time or) the biological stage T (in years, months, days, seconds, etc.) may be described in terms of the mentioned mood at (T-1) and the sum of the effects upon it (interactions c(I,J;K,L) and the feeding terms F(I,J;T-1)) as;

$$C(I,J;T) = F(I,J;T-1) + (C(I,J;T-1) + \sum_{K,L}^{nn} c(I,J;K,L)C(K,L;T-1))/(\rho+1) \quad , \qquad (1)$$

where ρ is the number of the nn entries for (I,J) and the sum is over the nn entries. The range of the interactions and thus the number of the nn entries (ρ) in Eq. 1 may be selected arbitrarily and the connections c(I,J;K,L) may be uniform (same for all of the entries (I,J) and (K,L)) or non uniform (not same, similarly). Please note that the Eq. (1) is a “coupled map” for the real C(I,J;T); while for integer (in particular, binary) it would be called "cellular automata" (please see Appendix and the references mentioned there for cellular automata and further discussions about the related issues). Please note that the unit for T is arbitrary and c(I,J;K,L) involve the total interaction within one unit of time for T; i.e., between T and T-1. As a result c(I,J;K,L) may be taken bigger than unity and similarly for the other parameters, such as the feedings F(I,J;T) and so on. Hence, T may be small or big for the similar organization of the designs if the mentioned parameters are taken big or small, respectively.

Synchronous (parallel) updating is followed in (Eq. (1)) all of the applications of the model.

*Generation of the model signals*: Each entry of the lattice C(I,J,T) may be taken not only as a part of a spatial pattern but also a biological source producing some electromotive force or (current or) voltage difference (V) with respect to a ground value. In other words, the mentioned electrical properties may be product of the same (or similar) processes which govern the spatial formations. Thus, each column (or row) of the lattice C(I,J;T) may be taken as a time series for a pulse $V_I(t;T)$ (or $V_J(t;T)$) in an electrical signal produced by the biological units (channels) sited along the columns with I=1, 2, ..., N (and similarly for J for the row wise configuration) where t (J→t or I→t) is the time for the signal which emerges (in a column wise or row wise channel as described in the next coming equation; Eq. (2)) at the biological stage T. In this manner, the numerical values of the entries within each column (I) or row (J) of the lattice may be considered as a part of an output (beat) for the electrical signal with duration (time period) equals to N in arbitrary unit.

We take V(t;T)=C(I,J;T) for the signal which is composed of many pulses coming from different emitters (column channels) where t is an integer variable which increases by one:

$$t=(I-1)N + J \quad . \qquad (2)$$

In the Eq. (2), I and J increase in a nested manner; that is, J increases from 1 to N for I=1 and then I increases to 2 and J increases from 1 to N again for this new value for I. This goes on till I=N, J=N; thus t increases by one from 1 up to $N^2$; i.e., t=1,2, …, $N^2$.

Similarly, the initial conditions for C(I,J;T=0) could be treated as the initial signals V(t;T=0) = C(I,J;T=0) where t is the same as before.

The pulses may be similar or the same and thus the resulting signals may be periodic with the period (P; where P=N is in arbitrary unit of time). Secondly, the numerical expressions for the model signals (for a T) may be computed directly in terms of some secular equations (through the Eq. (1)) for non random parameters. If instead some random parameters are utilized for the initial excitations or the feeding terms, then the same map may be used for simulations, equally well.

The model may be applied to various biological (or non biological) examples with various assumptions and several parameters for the initial and boundary conditions, the connections and the feeding terms and so on. The next section is the applications of the model where the main aim is to model the heart or brain signals of human. Some electrical activities in several muscles of humans or various neurons of animals or some spatial formations are considered in Appendix.

It may be worth to underline that the initial conditions (signals; V(t;0)), the feeding terms (F(I,J;T)) and the number for the iterative tours (T) are important for the outcomes of the present model.

**3. Applications and results:** A square (NxN) lattice (C(I,J;T)) with I≤N and J≤N is taken to represent a group of biological elements in heart or brain (or several muscles or neurons in Appendix) at a temporal stage designated by T where T is an integer variable which increases by unity, i.e., T=1, 2, … . Furthermore, periodic or no boundary conditions may be assumed. A periodic boundary condition means in the first nn approximation that the first nn entries of an entry sited on a side (row wise or column wise) of the lattice, involve the nn entries from the opposite side (row wise or column wise) and ρ=4 in Eq. (1). On the other hand, no (or fixed) boundary condition means that no entry sited on a side (row wise or column wise) of the lattice, involves no nn entry from the opposite side (row wise or column wise). Secondly, within the first nn approach; ρ=5 for an entry in the bulk (i.e., K-1≤I≤K+1 and L-1≤J≤L+1), ρ=4 for an entry on a side (i.e., I=1 or I=N or J=1 or J=N but not on a corner) and ρ=3 for an entry on a corner (i.e., I=J=1 or I=J=N or I=1 with J=N or I=N with J=1).

*Connections*: The strengths for the row wise (horizontal; h) and column wise (vertical; v) connections may be different for non uniform connections:

c(I,J;K,L)=v if I=K and J=L-1 or J=L+1 (in the first nn approximation)
or c(I,J;K,L)=h if J=L and I=K-1 or I=K+1 (similarly) . (3)

It is clear that v=h for uniform connectivity and the evolution rates per T for the designs C(I,J;T) increase as connectivity strengths (within the present first nn approach) increase.

*The feeding terms*: The parameters F(I,J;T) in the Eq. (1) may be constant (independent of T or variable for various species or organs and F(I,J;T) may involve random or non random terms. If the feeding terms are taken as independent of time;

$$F(I,J;T)= B + R\lambda \quad , \qquad (4)$$

where B and R are some (positive for the activators or negative for the inhibitors) real numbers and $\lambda$ is a uniform real random number with $0\leq\lambda<1$.

In some cases the time dependent feedings may be used in the following form (please see Sec. 5.1 in Appendix);

$$F(I,J;T)= (B + R\lambda)/T \quad , \qquad (5)$$

where B and R are same as in Eq. (5) and F(I,J;T) saturates with T, i.e., $F(I,J;T)\rightarrow 0$ as $T\rightarrow\infty$.

In summary; for the temporal and biological formations, each column (or row) of the representing lattice is considered as a different signal emitter in an organ. The emitters interact with each other as described beforehand. The topography of the initial conditions may be considered as the variation (design) in voltage amplitudes in the initial signals.

*Initial signals*: We think that the initial bio-signals (excitations or triggers) are important for the organization of the outcomes. Secondly, they may be changing from one species to the other within various organs (human heart, human brain, several muscles or single neuron of human or various animals and so on). Thirdly, they may also be changing from time to time within the same (tissue or) organ of a human or animal (in several kinds of epileptic seizures, for example) due to some biological reasons. Several initial signals are treated differently for different cases in the following sections with the declared parameters in the Table I where a special section is devoted for the spiral waves with the related parameters depicted in the Table II. The parameters used in Appendix are given in the Table III.

Please note that, synchronous (parallel) updating is followed in (Eq. (1)) all of the applications presented here or in Appendix.

**Model action potential:** N=30 is taken (arbitrarily) for modeling a starting action potential (spike, burst) [5] where the initial signal is taken Gauss;

$$V(t;0)=A\exp(-(t-\tau)^2/\sigma) \qquad \text{if t is even (odd)}$$
$$\text{or} \quad V(t;0)=0 \qquad \text{if t is odd (even)} \quad . \qquad (6)$$

It is clear that the pulse defined in the Eq. (6) is for the first period of the signal (with $t\leq N=30$) where t is the same as in the Eq. (2) and the maximum (A) of the pulse occurs at $t=\tau$. Secondly, periodic boundaries, non uniform connectivity (with h=0, v=0.1 in (Eq. 3)) and constant feedings (with R=0 in (Eq. 4)) are assumed. The plots in the Figures 1 - 3 are for the time (T) evolutions of the initial excitations (Eq. (6)), where the outcome are some model spikes for small T ($\leq 4$, say). The Figures 4, 5 show some statistical data about the model signals given in the Figs. 1 – 3, where the voltage amplitudes disperse with increasing T and the contour plots in the Fig. 5 are the spatial designs (patterns) C(I,J;T) for various T which are declared within the related plots.

The initial pulse (Eq. (6)) may in fact be considered as the product of the given Gauss with an alternating wave, where the alternating (or step) wave is unit if t is (say) even or zero otherwise. A suitable exponentially increasing function or a linearly increasing (steep, with big slope) one after some big t (<30) may be substituted for Gauss (and so on) to obtain similar theoretical plots (not shown). But, if fixed boundaries are assumed (instead of the assumed periodic ones) then the fine structures about the minima for the spikes become flat (not shown).

Please note that no random parameters are used in the present application (Table I) and all of the plots in the Figs. 1-5 are computed (not simulated) in terms of some small T ($\leq$4, say) in the map (Eq. (1)). Please see the Figure 6 which is borrowed form [**6**] **(**open access) for the realistic action potential curves. The next section is the model heart or brain signals of human.

**Model human heart or brain signals:** The Figure 7 is the model signal with an initial excitation where each pulse is sum of two Gaussians with the same amplitude but different σ;

$$V(t;0)=A(-\exp((t-\tau_1)^2/\sigma_1) + \exp(-(t-\tau_2)^2/\sigma_2)) \text{ with } t\leq P=N=20 \quad . \tag{7}$$

The model parameters for the Figure 7 are given in the Table I where there is no random parameter and the design of each pulse and signal in the Fig. 7 may be computed in terms of simple algebra for T=2 (Eq. (1)). The Figure 8 is the schematic representation of the real and normal heart beats (ECG) which is borrowed from [6] and the Figures 9, 10 are the empirical data showing some heart illnesses called atrial fibrillation and slow heart rate or (sinus) bradycardia, which are borrowed from the mentioned source. The Figures 11 and 12 are the model signals where a random component (R=0.2) is incorporated in the feedings besides a non random component (B=5) or B=1 is used with zero random feedings, respectively. Please note that, if the signals coming from some randomly selected few emitters within the model signals of the Fig. 11 with 0«I«20 are omitted, then an outcome similar to the empirical plot in the Fig. 9 (upper) may be obtained (not shown).

The Figures 13 (a)-(d) exhibit the empirical data for human heart signals which are borrowed from [2]. The pulse waveforms are recorded for the following physiological conditions; normal (a), quasi-stable (b), unstable (c) and post-operative (stable) (d) cases. One may notice by inspection of the Figures 13 (a)-(d) that the plots are composed of somewhat similar (repeated) pulses which are more or less periodic, except the plot in the Fig. 13 (c) which is nearly a random signal. Secondly, the wave forms of the pulses in the Fig. 13. (a) are approximately Gauss, if the kinks for the extremes are disregarded (which may also be considered as Gauss with small mean variance). Hence, a periodic initial signal which is composed of Gauss pulses may be assumed;

$$V(t;0)=A\exp(-(t-\tau)^2/\sigma) \text{ with } t\leq P=N=20 \quad . \tag{8}$$

It is clear that the pulse given in the Eq. (8) is for the first period of the signal, where t is the same as in the Eq. (2) and the maximum of the pulse (A) occurs at t=τ.

The initial signal evolves in time as described in the Eq. (1) with the feeding terms (Eq. (3)) and the first nn interactions (connections) where a uniform coupling is assumed for the connections with v=h=0.5 (Eq. (3)).

The Figures 14 (a)-(d) are the model signals after following the mentioned assumptions and the model parameters. The random component (R) for the individual feeding terms (Eq. (3)) is zero for the Figs. 14 (a) and (d); hence, the topography of each pulse within the theoretical plots of the Figs. 14 (a) and (d) and these of the corresponding signals may be computed (not simulated) numerically in terms of few products or sums of numbers (or Gauss functions for analytical derivation) with T=9 or T=21 (Eq. (1)), respectively. The figures 14 (b)-(c) are simulations for T=21, where R is the biggest for the Fig. 14 (c).

Please note that T=21 for the Figs. 14 (b)-(d) whereas T=9 for the Fig. 14 (a) and the interaction strength for the connections is equal to 1/2 in all. The Figure 15 is same as the Fig. 14 (a) but for the uniform connections with smaller strength (equals to 0.15); as a result, the number of the iterative tours is big (T=21).

It may be worth to mention that the similar plots to the ones in the Figs. 1-14 may be obtained under different assumptions for the initial signal (conditions), boundary conditions, connections, feeding terms and number for the iterative tours (T) and so on. For example, a similar plot to the one in the Fig. 14 (a) may be obtained with a sinusoidal initial signal (not shown). (Please see also the Figure 23 (b).)

Now, let us consider random initial signals;

$$V(t;0)=A\lambda(t) \quad , \tag{9}$$

where $\lambda(t)$ means that a different random real number between zero and one is tried for each t and A is a real number for the amplitudes.

The Figures 16 (a) and (b) show model signals for various T which increases by one from the bottom (T=0) to the top (T=10) and the initial signals (T=0) are random with unit amplitudes Eq. (9) in both of the plots; but, the feedings are different. In the Figure 16 (a) the feeding terms involve no random component whereas B=R=1 in the Figure 16 (b). The Figures 17 and 18 are various EEG for absence seizure, which are modified from the Figs. 2 and 5 in [7] and [2], respectively. The Figure 19 is for some model signals where a periodic and rectangular (steps or wells) initial wave is used;

$$\begin{aligned} V(t;0) &= 0, \quad \text{if } 0<t\leq S1 \\ \text{or} \quad &= A, \quad \text{if } S1\leq t\leq S2 \\ \text{or} \quad &= 0, \quad \text{if } S1\leq t\leq P \quad . \end{aligned} \tag{10}$$

In Eq. (10), S1 and S2 are some integer parameters; A is the amplitude (height of a step) and P is the period of the initial signal with P=N=20 and t is same as in the Eq. (2) and T=3. Hence, the sequence of the pulses in the Fig. 19 is computed numerically out of the coupled algebraic equations (Eq. (1)) for T=3.

The Figure 20 depicts some empirical data for childhood absence epilepsy and the Figures 21 (a)-(h) are some model signals where random initial waves (Eq. (9)) and non uniform connectivity (h=0, v=0.1) are used with zero feedings (N=100).

**Spiral waves as the initial excitations for model signals of human heart or brain:** There are many empirical evidences about the propagation of spiral waves in the tissues of various organs [9]. In this section we treat the spiral waves as the initial signals for various model bio-signals.

Spirals S(t) are the trajectories of the points $(S(r,\theta))$ where the radial distance (r) and the angle ($\theta$, with respect the horizontal axis) of a point on the spiral $(S(r(t'),\theta(t'))$ change with time;

$$S(r(t'),\theta(t'))=S(r_0 + Vt', \theta_0 + Wt') \quad . \tag{16}$$

In the Eq. (16) $r_0$, V, $\theta_0$, W stand for the origin, the radial speed, phase and the angular speed, respectively and the spiral propagates clock wise for W<0. Please note that the time parameter t' in the Eq. (16) is not same as in the Eq. (2); it is the time for the propagation of the spirals. In the applications here, we assume that t' increases by unity in each iterative tour; i.e., $t'\rightarrow t'+1$ if $T\rightarrow T+1$ where the related parameters are selected accordingly. Furthermore, the spiral waves are taken as emerging at the center of the representing lattice with an odd number for N. (It is clear that $t'\rightarrow t'+\gamma$ if $T\rightarrow T+1$ could be selected for any $\gamma$ and then the corresponding parameters should be changed accordingly to obtain similar outcome.) Hence;

$C(I,J;0)=1$ if $I\text{-}I'=S(r(t'),\theta(t'))=r(t')\cos(\theta_0 + Wt')$ and $J\text{-}J'=r(t')\sin(\theta_0 + Wt')$
or $C(I,J;0)=0$ otherwise , (17)

where $I'=J'=((N-1)/2)+1$ (and N is odd). Elliptical (with the eccentricity different than unity) spiral waves are disregarded. The following parameters are selected for all of the applications here: $r_0=0$, V=0.2, $\theta_0=0$ and W=-0.1 with arbitrary units and N=99. One may have different number of total points (M) on the initial topologies with the same parameters defining the spiral waves in different (NxN) lattices, since the spirals are circular but the representing lattice is not. Please note that the parameter M (Table II) counts the number of the entries (of the lattice) occupied by the initial spiral at T=0. V(t;0)=C(I,J;0) is taken for the initial signal where t is the same as in the Eq. (2).

The spiral wave with the mentioned parameters is shown in the Figure 22 (a) (left) and several model signals emerging out of the propagation of the given spiral wave are shown in the Figures 22 (b) – (c) for various t and T where the variation in the topography (design, similarity) of the beats is clear which is called spatiotemporal change (or variation) in the literature.

Now it is assumed that the amplitudes of the initial spiral waves may be multiplied with some Gauss which means that the emitters are not uniform and they function with different amplitudes; i.e.,

$C(I,J;0)=40\exp(-(t-\tau)^2/\sigma)$ if $I\text{-}I'= r(t)\cos(\theta_0+Wt)$ and $J\text{-}J'=r(t)\sin(\theta_0+Wt)$
or $C(I,J;0)=0$ otherwise , (18)

where A is the amplitude and the maximum of the signal is for $t=\tau= -40$ and with $\sigma=1000$.

The Figures 23 (a) – (b) depict several model signals with initial spiral waves defined in the previous paragraph. Please note that the number of the entries with non zero amplitudes (bigger than 0.01, more properly) is same as before (=252) but the related amplitudes are different (Eq. (18) and Fig. 23 (a), right). In the Fig. 23 (b) V(t,91) is shown within various epochs where the plot for the time domain $4455 \leq t \leq 5445$ and (down and right) may be considered as similar to the empirical data in the Fig. 13 (a).

The spirals may have many arms and we consider here a 4-armed one where each arm (n+1) has a phase difference $\theta_{0,n+1} - \theta_{0,n}$ with respect to the arm (n);

$$\theta_{0,n+1} - \theta_{0,n} = 2\pi/5 \text{ with } 1\leq n\leq 5 \qquad (19)$$

and the initial configuration of the representing lattice becomes

$C(I,J;0)=40\exp(-(t-\tau)^2/\sigma)$ if $I\text{-}I'= r(t)\cos(\theta_{0,n}+Wt)$ and $J\text{-}J'=r(t)\sin(\theta_{0,n}+Wt)$
or $C(I,J;0)=0$ otherwise . (20)

Some model signals with initial 4-armed spiral waves are displayed in the right inset of the Figure 24 (a) where the amplitudes are multiplied by a Gauss factor (Eq. (19)); please see the Eqs. (19) and (20) and the related text. The number of the entries covered by the 4-armed spiral with non zero amplitudes (bigger than 0.01, more properly) is little bigger than 1000 here. Please note that the time domains ($4455 \leq t \leq 5445$) are common in the plots: Fig. 24 (**a**) for T=0 and (**b**) T=91 where the insets show the distributions of the number of the initial amplitudes over the mentioned amplitudes (histogram, left) and the initial amplitudes over the entries (I,J) of the model organ (right), respectively.

**4. Discussions and Conclusion:** It may be observed within the given model signals that the same (or similar) outcomes may be obtained under different assumptions for the initial signal, connectivity, feeding term or number of the iteration tours (T) and so on. Yet, some hierarchy (regularity or order) occurs within the outcomes for the model signals. For example, as T increases ($T \rightarrow \infty$) the results become periodic whatsoever the topography of the initial signal or the assumption for the boundary or the parameter for the feedings are. Secondly, the evolved signals come out as concave up (concave down, convex) for the activators (inhibitors) with big T.

Some terms for the exponential decays in the voltages at the sites (I,J) could also be incorporated within the present model. Yet, the effect of the mentioned decays with T may be absorbed within many of the present model parameters declared in the Eq. (1). Consider

$$C(I,J;T)=F(I,J;T-1)+\exp(-\alpha)(C(I,J;T-1)+\Sigma_{K,L}^{nn} c(I,J;K,L)C(K,L;T-1))/(\rho+1) \quad , \quad (21)$$

where the exponential term ($\exp(-\alpha)$) is a simplification for $\exp(-\alpha((T-1)-T))$; i.e., decays.

One may organize the Eq. (21) in the following way;

$$C(I,J;T)=F(I,J;T-1)+[-1+1+\exp(-\alpha)](C(I,J;T-1)+\Sigma_{K,L}^{nn} c(I,J;K,L)C(K,L;T-1))/(\rho+1) \, , \quad (22)$$

$$C(I,J;T)=F(I,J;T-1)+ (C(I,J;T-1)+\Sigma_{K,L}^{nn} c(I,J;K,L)C(K,L;T-1))/(\rho+1) + \{[-1+\exp(-\alpha)](C(I,J;T-1)\} /(\rho+1)+\Sigma_{K,L}^{nn} [-1+\exp(-\alpha)]c(I,J;K,L)C(K,L;T-1))/(\rho+1) \, , \quad (23)$$

$$C(I,J;T)=\{F(I,J;T-1)+[(-1+\exp(-\alpha))(C(I,J;T-1)/(\rho+1)]\} +(C(I,J;T-1)+\Sigma_{K,L}^{nn} c(I,J;K,L)C(K,L;T-1))/(\rho+1) + +\Sigma_{K,L}^{nn} [-1+\exp(-\alpha)]c(I,J;K,L)C(K,L;T-1))/(\rho+1) \quad . \quad (24)$$

Now, one may rename the terms or the expressions in the Eq. (24)

$$C(I,J;T)=F'(I,J;T-1)+(C(I,J;T-1)+\Sigma_{K,L}^{nn} c'(I,J;K,L)C(K,L;T-1))/(\rho+1) \quad , \quad (25)$$

where it is clear that $F'(I,J;T-1)=F(I,J;T-1)+[(-1+\exp(-\alpha))(C(I,J;T-1)/(\rho+1)]$ and $c'(I,J;K,L)= \exp(-\alpha)]c(I,J;K,L)$. So there is no need to consider the relaxation constant (or the related time constant, $\alpha$) for the voltages or the currents within the model cells explicitly.

## 5. APPENDIX

A simple application of the model for spatial formations in biology may be solar darkening of human skin where the accumulation of the resistive pigments in the skin cells saturates with time although exposures of the human to the Sun may continue. Hence, $C(I,J;T=0)=0$ for all I and J and $F(I,J;T) = (B + R\lambda)/T$ (Eq. (5)) may be taken (with some non zero parameters for B or R) for the initial conditions and the feedings, respectively. Furthermore, periodic boundary conditions with zero connections may be assumed. As a result, C(I,J;T) saturates with some fluctuations as T increases (not shown).

The cellular automata is considered as the second application of the present model for spatial but non biological formations, in the section 5.1; namely, freezing of a liquid (or melting of a solid) within a pot. The biological and temporal formations are further treated in the section 5.2 for the electrical activities in various muscles or single neurons of human or several animals. The first nn approximation is followed for all of the mentioned cases with different assumptions for the initial and boundary conditions, connections and various parameters and so on.

**5. 1 Spatial and non biological formations; cellular automata for freezing (or melting):** A group of interacting agents (say, water molecules which heat up or cool down in time) may be represented by a two dimensional lattice (mesh), where the interactions (i.e., heat) may be averaged between the neighbor molecules (assuming that the individual heat capacities are constant) at each time T. It is known that if the temperature of the liquid is equal to the freezing (or melting) temperature, then some portion (some group of molecules) of the liquid may be found as frozen. Secondly, the material continues to freeze or melt more and more depending upon whether the heat is supplied or extracted from the medium, at the mentioned temperature which remains constant (due to the so called latent heat) during the freezing (or melting). The important point is that, there may be a clear distinction between the frozen group of molecules and the others. Hence, the frozen (not frozen) portion of the liquid (say, at the surface), i.e., the related entries of the representative lattice may be designated by the number zero (one), or vice versa. And accordingly the supplied (or the extracted) heat may be designated by a positive (or negative) parameter and so on.

Here, a square (100x100) lattice is taken to represent some liquid (or solid) within a pot. Secondly, no boundary condition is assumed (Section 3) and we start with a random initial condition, where C(I,J;T=0) (Eq. (1)) may be zero or one according to the following rule;

$$C(I,J;0)=0 \quad \text{if } 0\leq\lambda(I,J)<0.5$$
$$\text{or} \quad C(I,J;0)=1 \quad \text{if } 0.5\leq\lambda(I,J)<1.0 \quad , \qquad \text{(A1)}$$

i.e., in terms of rounding the real numbers to zero or one. [12] In the equation (A1) λ(I,J) means that a different uniform and real random number is tried for each (I,J); thus, we initially have about the same amount of liquid or solid material within the pot.

For the evolution, we follow the coupled map of the Eq. (1) with integer numbers (zero or one) for C(I,J;T) where the real numbers for C(I,J;T) with 0<T are changed with zero or one by rounding as in the Eq. (A1), i.e., C(I,J;0)→C(I,J;T). Secondly, uniform connections with unit strength are assumed. Thirdly, the feeding terms are taken as time dependent (Eq. (5)).

The Figures 25 (a)-(c) are the results of the present model for the spatial formations for freezing (or melting) of a liquid (solid) in terms of the cellular automata with the assumptions mentioned in this section and the parameters given in the Table II. The Fig. 25 (a) is for the distribution of the parameters for C(I,J;T=2) over the entries (I,J). The Figure 25 (b) depicts the distribution of the number of the amplitudes (C(I,J;2)) over the mentioned amplitudes where the vertical axis is logarithmic. The Figure 25 (c) is the variation of the average of the (integer) amplitudes for C(I,J;T) (0<T) with time where the average (avg(T)) is defined in the usual manner; i.e.,

$$avg(T)=\sum_I^N\sum_J^N C(I,J;T)/N^2 \quad . \qquad \text{(A2)}$$

Please note that avg(T) (Eq. (12)) may be considered as the total extracted (or supplied) heat from (to) the medium (surface) in arbitrary unit, which saturates in time (T). Secondly, the spatial unit in the Fig. 25 (a) is square whereas the groups of molecules are known to have non square shapes. Yet, the present application may be considered as an approximation for the discussed situation(s) and it is clearly suitable for the formation of labyrinths where finding continuous paths may become easier or harder as T increases, depending upon which color (for the zeros or the unities) is taken to stand for the barriers.

**5.2 Modeling electrical activities in several organs:** Bio-signals coming out of hand muscles of human, single neurons of various animals are modeled in this section where the periodic boundary conditions are assumed for a square (NxN) lattice. The related parameters are given in the Table III for all of the applications treated in this section. Please note that no random parameters are utilized for the plotted model signals here and hence they are computed (not simulated) in terms of various secular equations (Eq. (1)).

**5.2.a Model electromyograms (EMG):** The model signals in the Figure 26 are obtained with the following initial signals:

$$V(t;0)=A \quad \text{if } t \text{ is even (odd)}$$
$$\text{or} \quad V(t;0)=0 \quad \text{if } t \text{ is odd (even)} \ . \qquad \text{(A3)}$$

It is clear that the pulse defined in the Eq. (A3) is for the first period of the signal (with $t \leq N=30$) where t is the same as in the Eq. (2) and A is the amplitude. Secondly, some non uniform connections are used and the feedings are time independent (Eq. (4)) with B=0.1 and R=0. Hence, the plot is computed in terms of 9 coupled secular equations (Eq.(1)). The Figure 27 is the empirical EMG (modified from [8]). The Figure 28 stresses the meaning (function or role of the) averaging process in the present model (Eq. (1) where the mean deviation of the voltage distributions about the average of the initial voltages ($\approx 150=(0 + 300)/2$) decreases with increasing T.

**5.2.b Model single neuron signals for various animals:** The Figure 29 depicts several empirical data for (single) neuron membrane voltage activities of various animals (modified from the Figure 3 in [10] which is modified from [11]): (a) Lobster pyloric neuron; (b) guinea pig inferior olivary neuron; (c) sepia giant axon and (d) mouse neocortical pyramidal neuron. The Eq. (6) is used for the excitations of the given model signals in the Figures 30 (a)-(b) and the following equations are used to generate the model initial signals for the plots in the Figures 30 (c) and (d), respectively:

$$V(t;0)=A\exp(-(t-\tau)^2/\sigma)\sin(t/3) \ , \qquad \text{(A4)}$$

where A is the amplitude and t, $\tau$ and $\sigma$ are as before (Eq. (6)) and

$$V(t;0)=A\exp(-(t-\tau)^2/\sigma) \quad \text{if } t \text{ is even (odd) and } S1 \leq t \leq S2$$
$$\text{or} \quad V(t;0)=\exp(-(t-\tau)^2/\sigma) \ \text{ otherwise} \ , \qquad \text{(A5)}$$

where all of the model emitters give out similar waves but few neighbors (between S1 and S2; S1~S2) with different amplitude.

**TABLES**

| explanation | N | initial wave | $1^{st}$ par. | $2^{nd}$ par. | $3^{th}$ par. | h | v | B | R | T |
|---|---|---|---|---|---|---|---|---|---|---|
| Figs. 1-5 (spike) | 30 | Eq. (6) | A = 3 0 0 | $\tau$=20 | $\sigma$=20 | 0.0 | 0.1 | -10 | 0.0 | various |
| Fig. 7 (heart) | 20 | Eq. (7) | A = 1 . 0 | $\tau_1$=7; $\tau_2$=16 | $\sigma_1$=20; $\sigma_2$=7 | 0.5 | 0.5 | 3.5 | 0.0 | 2 |
| Fig. 11 (heart) | 20 | same | same | same | Same | 0.5 | 0.5 | 5.0 | 0.2 | 4 |
| Fig. 12 (heart) | 20 | same | same | same | Same | 0.5 | 0.5 | 1.0 | 0.0 | 2 |
| Fig. 14 (a) (heart) | 20 | Eq. (8) | A = 3 . 0 | $\tau$=7 | $\sigma$=0.5 | 0.5 | 0.5 | 3.5 | 0.0 | 9 |
| Fig. 14 (b) (heart) | 20 | same | same | same | same | 0.5 | 0.5 | 3.5 | 0.3 | 21 |
| Fig. 14 (c) (heart) | 20 | same | same | same | same | 0.5 | 0.5 | 3.5 | 1.2 | 21 |
| Fig. 14 (d) (heart) | 20 | same | same | same | same | 0.5 | 0.5 | 3.5 | 0.0 | 21 |
| Fig. 15 (heart) | 20 | same | same | same | same | 0.15 | 0.15 | 3.5 | 0.0 | 21 |
| Fig. 16 (a) | 20 | Eq. (9) | A = 1 . 0 | none | none | 0.5 | 0.5 | 1.0 | 0.0 | 0-10 |
| Fig. 16 (b) | 20 | same | same | same | same | 0.5 | 0.5 | 1.0 | 1.0 | 0-10 |
| Fig. 19 | 20 | Eq. (10) | A = 1 . 0 | S 1 = 4 | S 2 = 1 8 | 0.5 | 0.5 | 2.0 | 0.0 | 3 |

**Table I** The model parameters are given in arbitrary units for the ones within the thick box at the middle which are for the initial signals. The other parameters are given column wise in the same but in arbitrary units.

| explanation | Eq. | M | h | v | B | R | T |
|---|---|---|---|---|---|---|---|
| Fig. 22 (a)-(c) | 17 | 252 | 3 | 3 | 1 | 0 | 11,31,91 |
| Fig. 23 (a)-(b) | 18 | 252 | 3 | 3 | 1 | 0 | 91 |
| Fig. 24 (a)-(b) | 19,20 | ~1000 | 3 | 3 | 1 | 0 | 0, 91 |

**Table II** The parameters for the model signals which evolve from the initial spiral waves where the parameters (please see the Eqs. (17)-(20) and the related text) are given column wise in the same but in arbitrary units.

| explanation | N | initial design | $1^{st}$ par. | $2^{nd}$ par. | $3^{th}$ par. | h | v | B | R | T |
|---|---|---|---|---|---|---|---|---|---|---|
| Fig. 25 | 100 | Eq. (A1) | none | none | None | 1 | 1 | 0.03 | 0 | 2 |
| Fig. 26 | 30 | Eq. (A3) | A=300 | none | None | 0 | 0.1 | 0.1 | 0 | 9 |
| Fig. 30 (a) | 30 | Eq. (6) | A=300 | $\tau$ = 0 | $\sigma$=100 | 0 | 0.1 | -10 | 0 | 8 |
| Fig. 30 (b) | 40 | Eq. (6) | A=10 | $\tau$ = 0 | $\sigma$=100 | 0.01 | 0.01 | 0.05 | 0 | 91 |
| Fig. 30 (c) | 100 | Eq.(A4) | A=5 | $\tau$ = 0 | $\sigma$=200 | 0.01 | 0.01 | 0.05 | 0 | 11 |
| Fig. 30 (d) | 40 | Eq.(A5) | A=10 | $\tau$ = - 3 | $\sigma$=150 | 0.01 | 0.01 | 0.05 | 0 | 11 |

**Table III** The model parameters are given in arbitrary units for the ones within the thick box at the middle which are for the initial signals (or the initial pattern; the first line which is for cellular automata). The other parameters are given column wise in the same but in arbitrary units.

## FIGURES

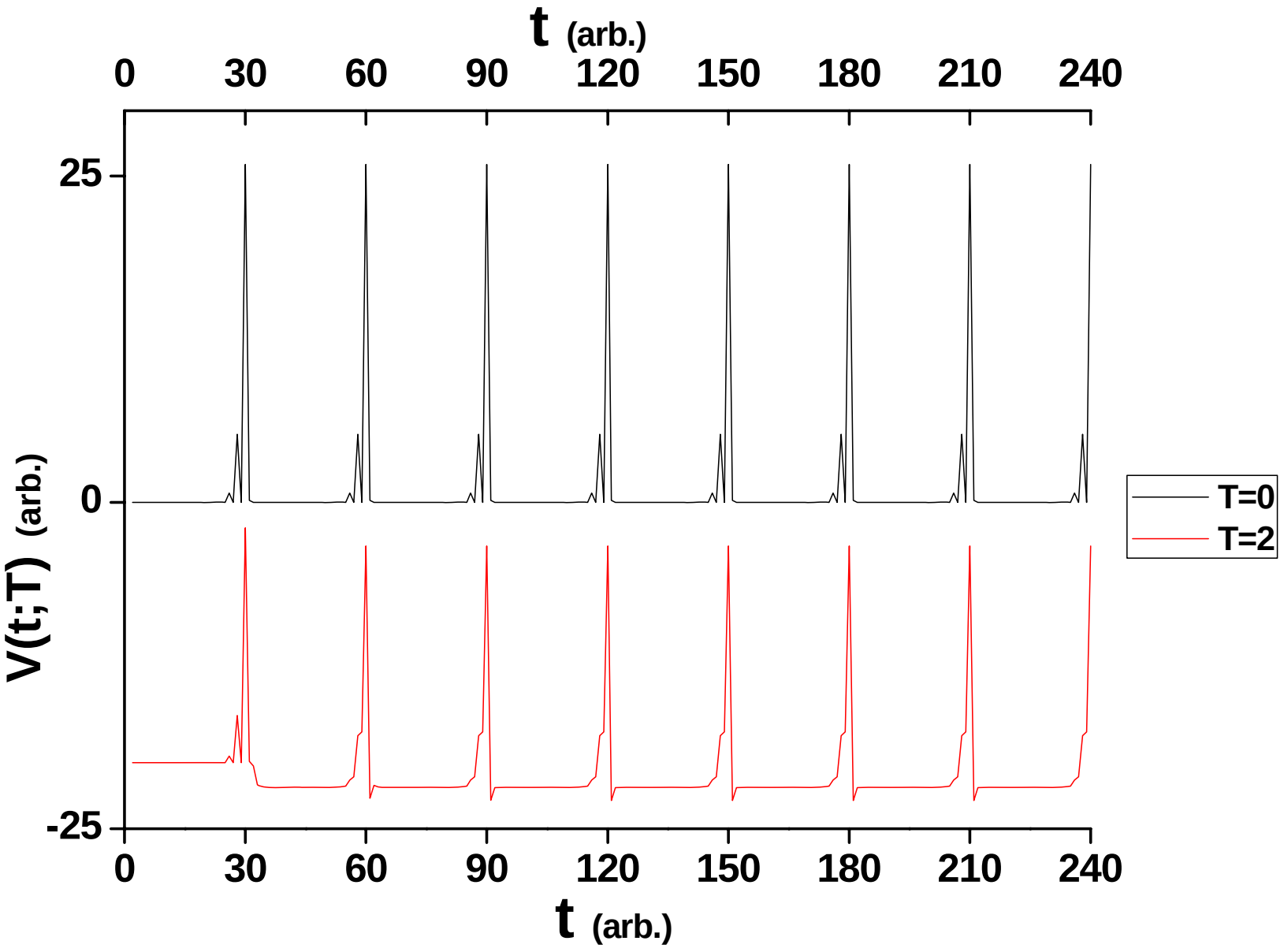

**Figure 1** Generation of model spikes for T=2 (bottom) out of initial pulses (top; C(I,J;T=0) in Eq.(1) which are defined in Eq. (6) in terms of V(t;0)) within a (30x30) representative lattice with the parameters given in the Table I and t≤900. Only the first eight (periods and) spikes are shown for clarity.

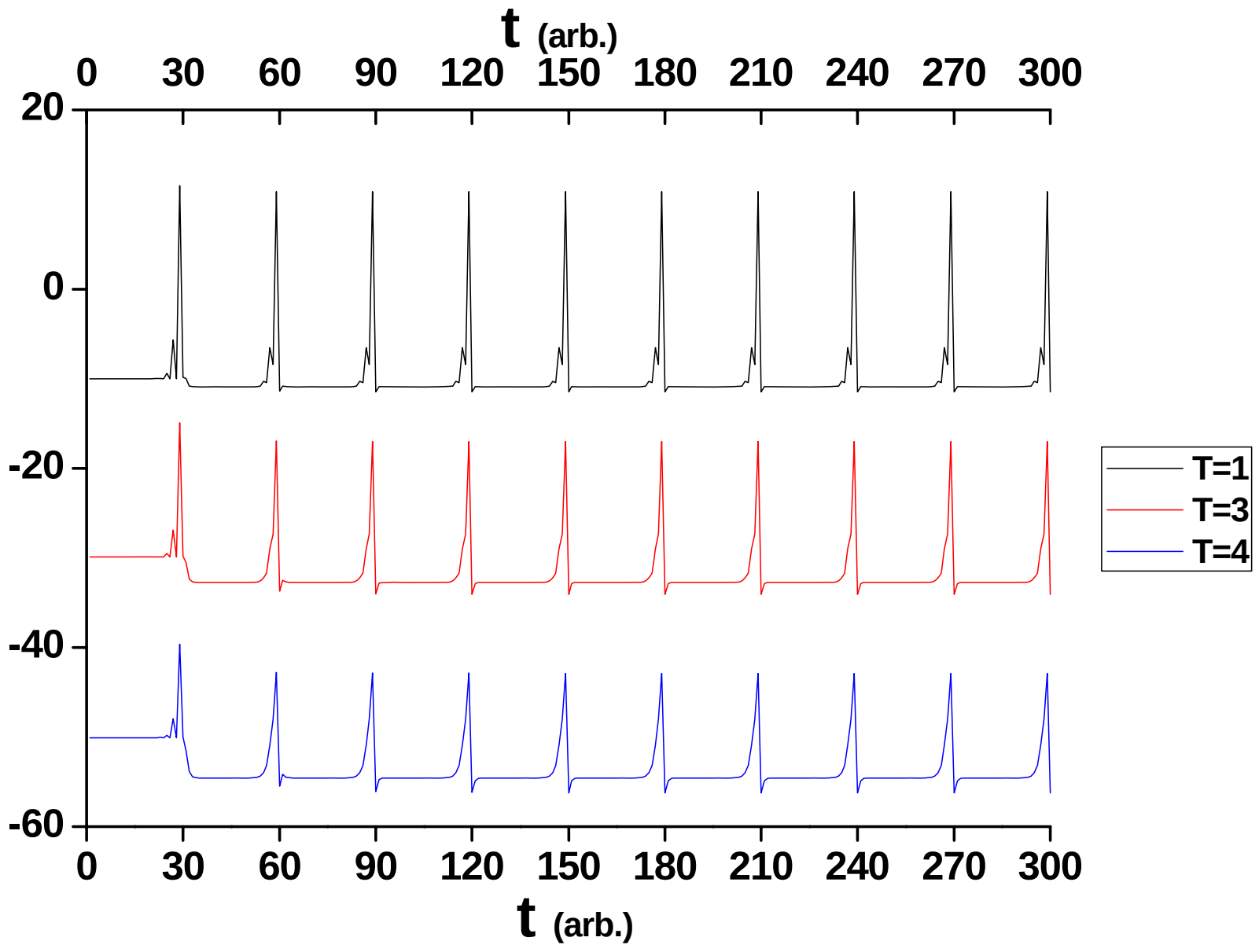

**Figure 2** Same as Fig. 1, but for T=1 (top), T=3 (middle) and T=4 (bottom). Only the first ten spikes are shown for clarity.

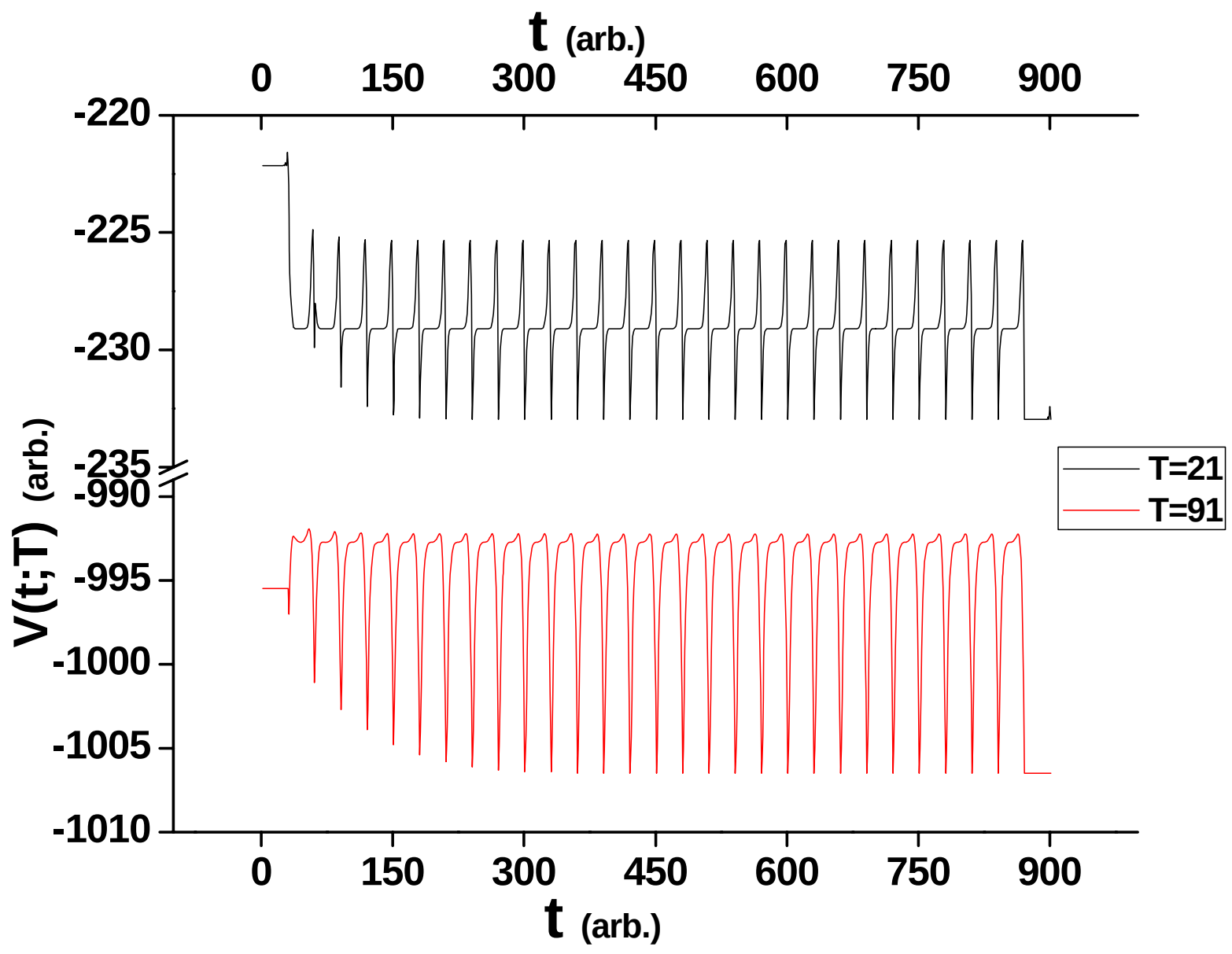

**Figure 3** Same as Fig. 1, but for T=21 (top) and T=91 (bottom), where all of the model emissions for a given T are shown.

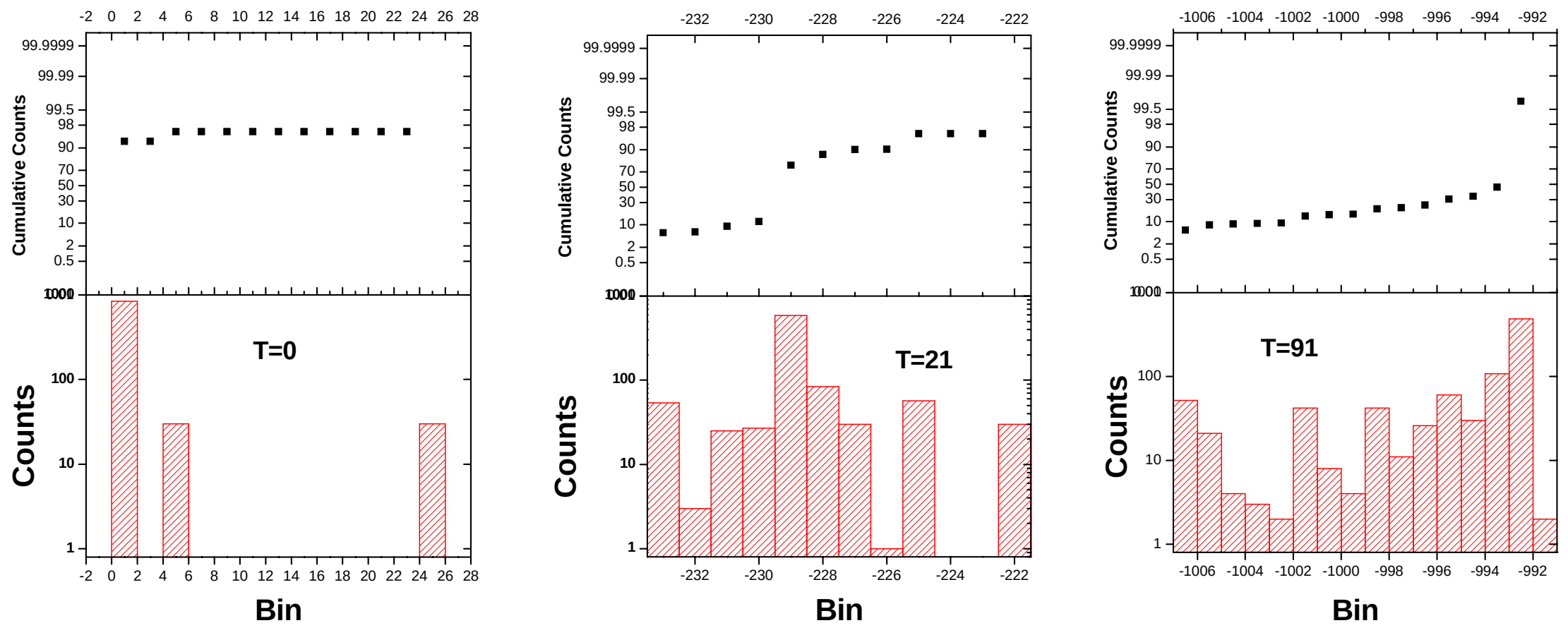

**Figure 4** The distribution of the numbers (counts) of the amplitudes of the voltages V(t;T) over V(t;T) (binned); simple distribution (histogram, below) and cumulative one (probability, dots, up) for T=0, 21 and 91 (from left to right) and t≤900.

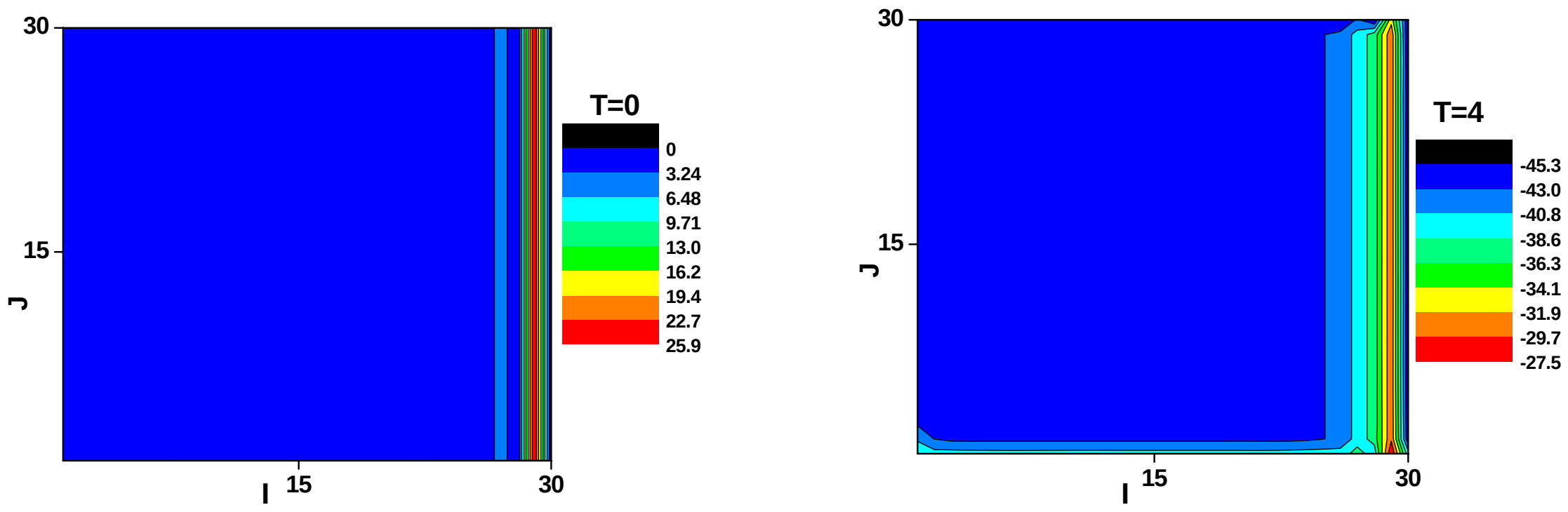

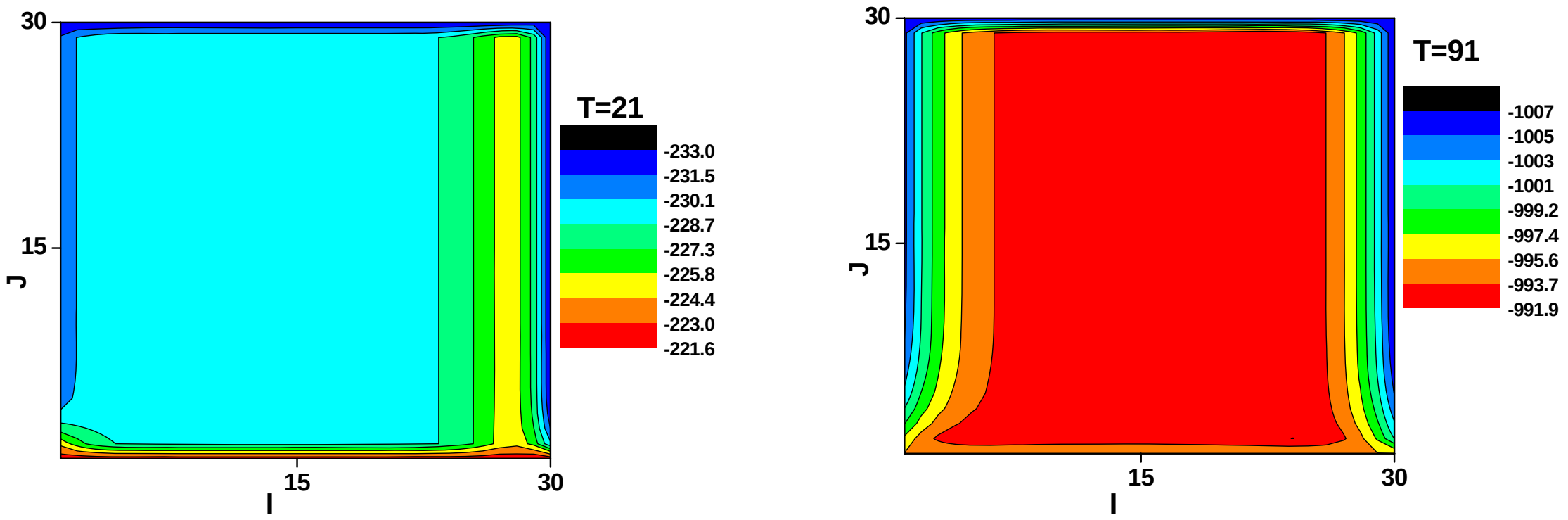

**Figure 5** The distributions of the voltage amplitudes (or the spatial patterns; C(I,J;T)) over (I,J) for T=0 and 4 (up row), 21 and 91 (bottom row) and I≤N=30 and J≤N=30.

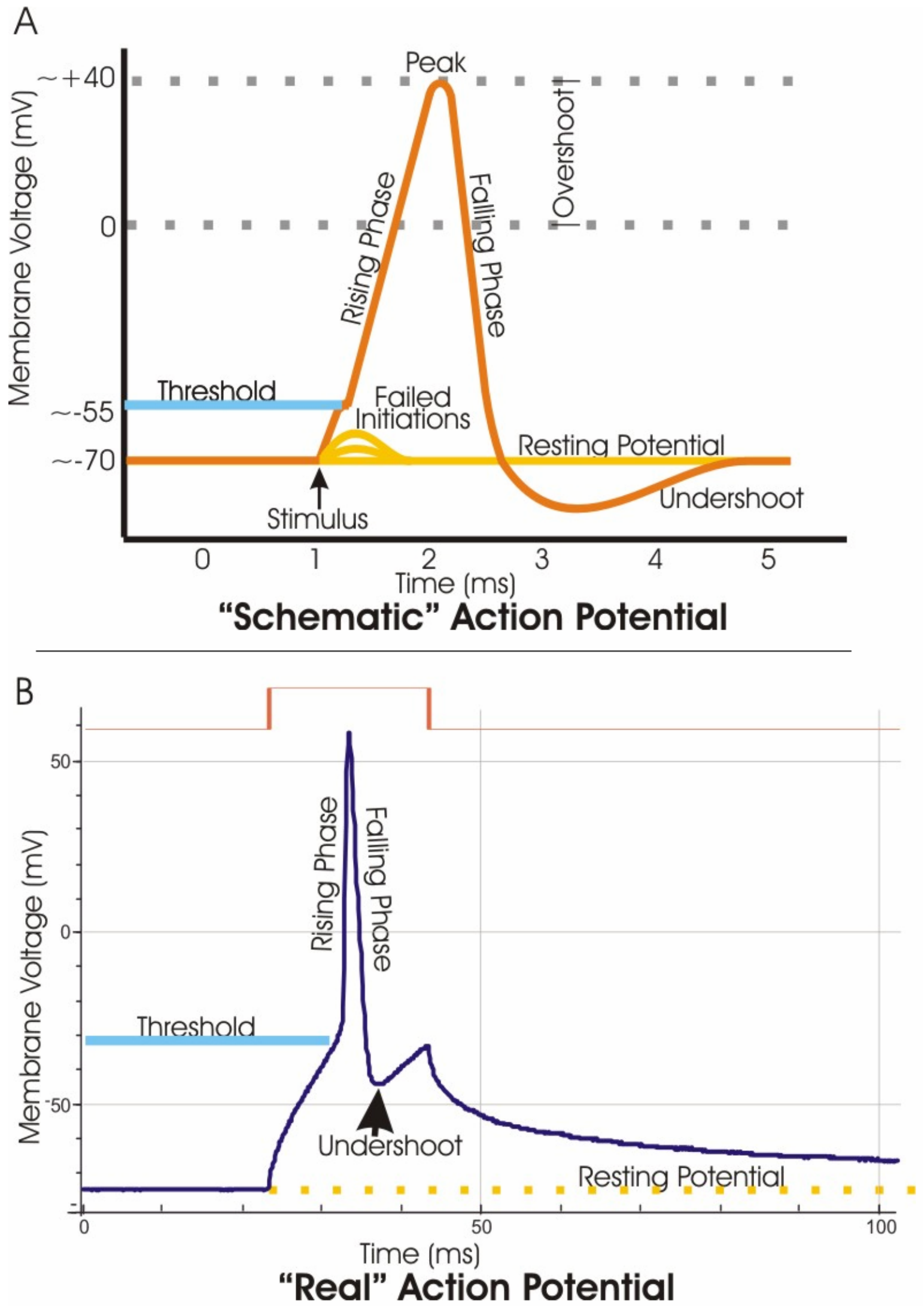

**Figure 6** Schematic and real action potential curves borrowed from [**6**] which is open access where some technical terms and symbols are depicted.

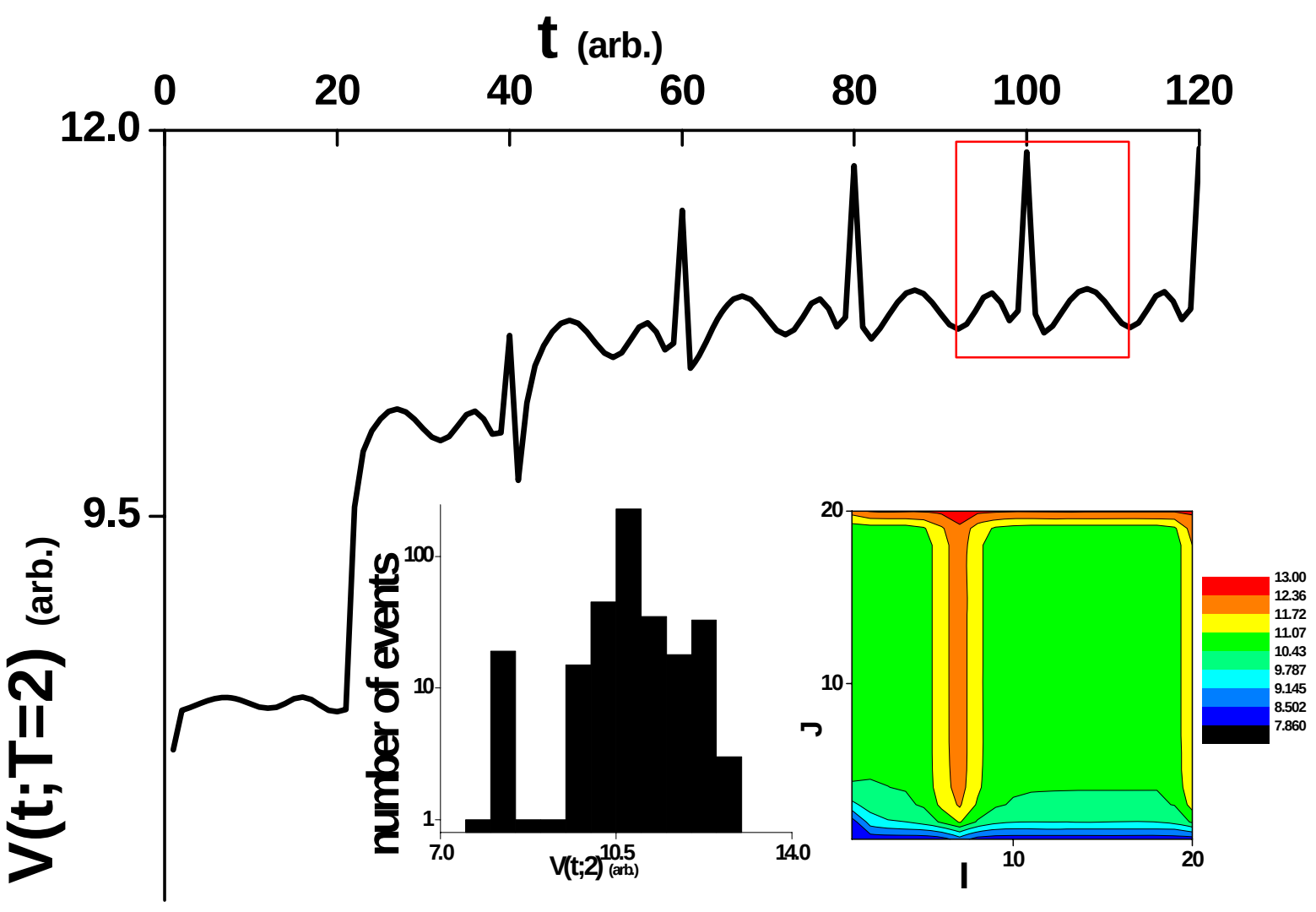

**Figure 7** Model signals generated for (20x20) representative lattice where sum of two Gauss pulses are used for the initial signal; Eq. (4) and Table I. The portion of the signal shown in the box (red) has 20 units of time for t as the time domain. The train of signals is (not simulated but) computed in terms of 2 secular equations (Eq. (1)). The insets show the distributions of the number of the model voltage amplitudes V(t;2) over the voltages (the histogram at left, where the vertical axis is logarithmic) and the same amplitudes for C(I,J;2) over the entries (I,J) of the lattice (at right), respectively.

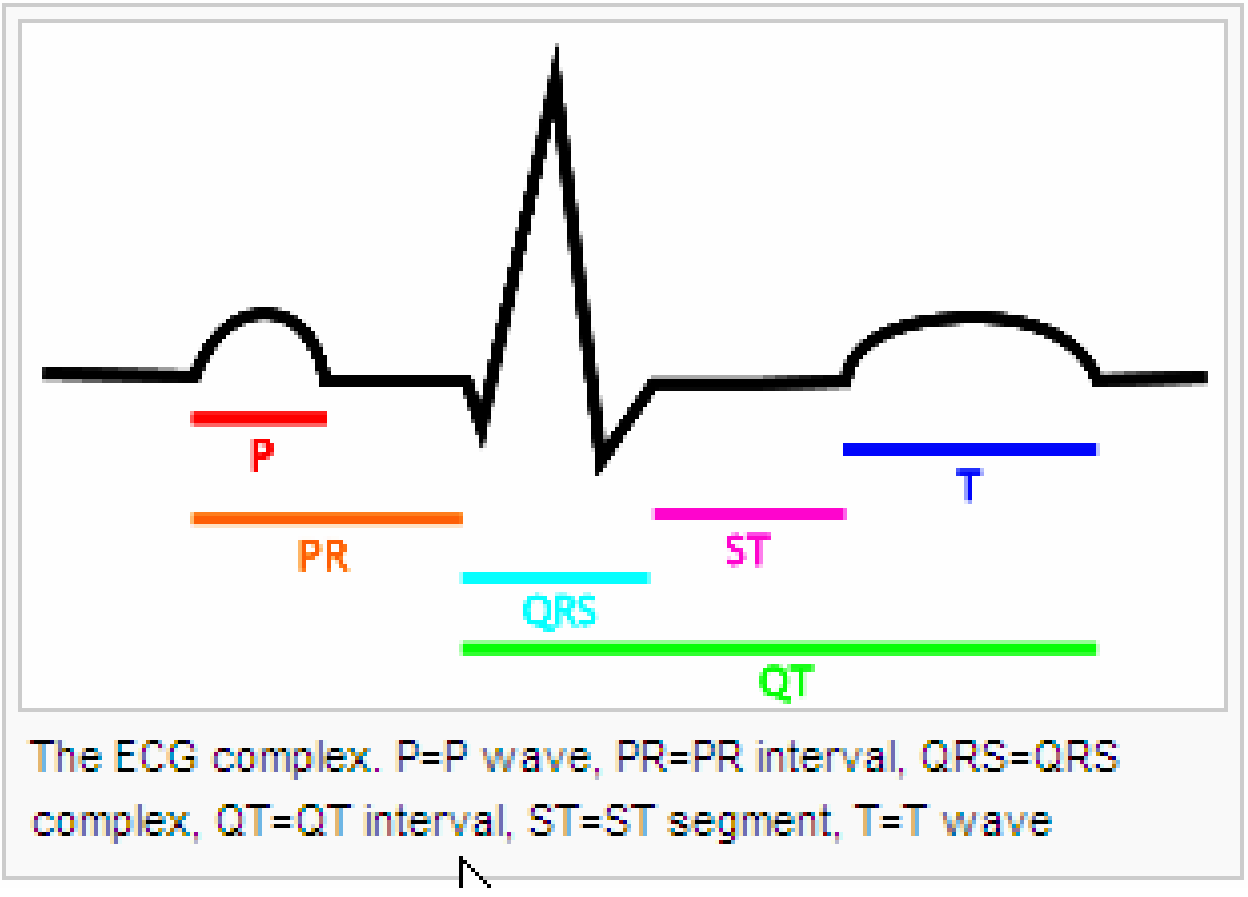

**Figure 8** Schematic representation of a pulse for the normal human heart where some technical terms and symbols are depicted. The figure is borrowed from open access [6].

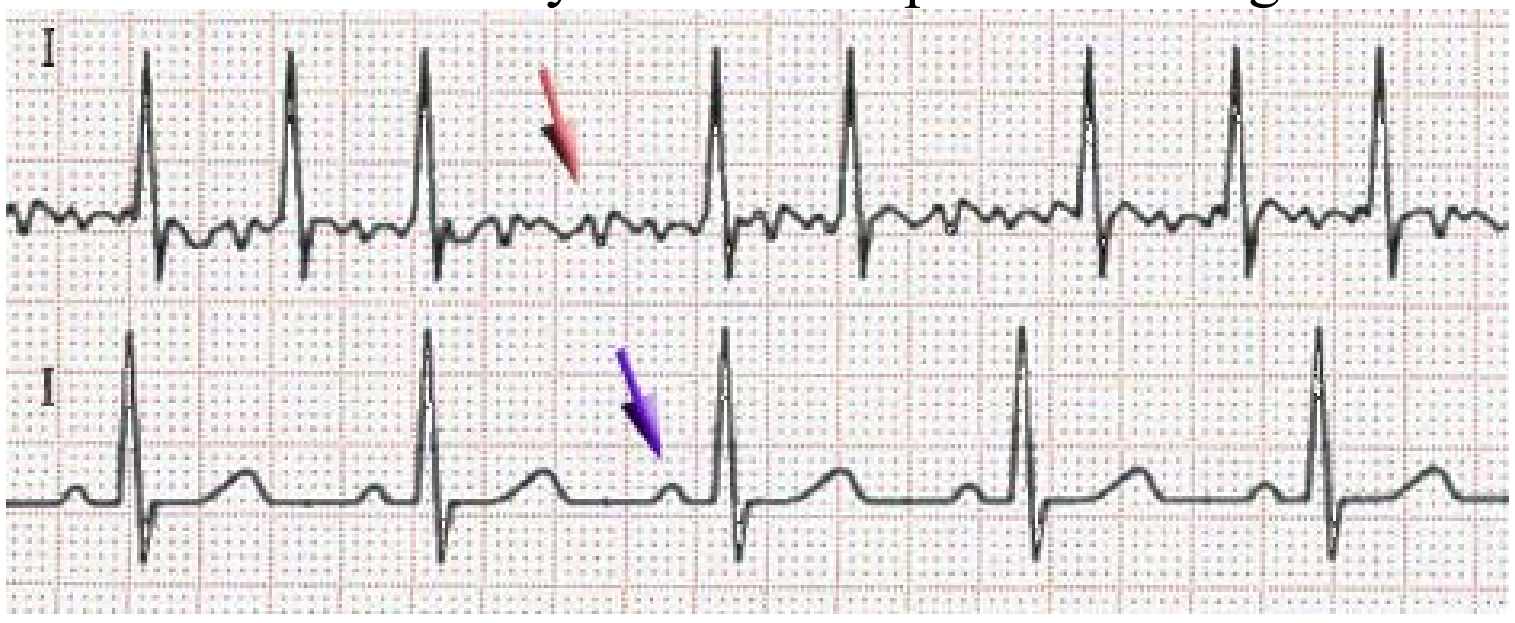

**Figure 9** Empirical data (ECG) of atrial fibrillation (top) and sinus rhythm (bottom). The arrow indicates a P wave, which is lost in atrial fibrillation. The plots are borrowed from [6] which is open access.

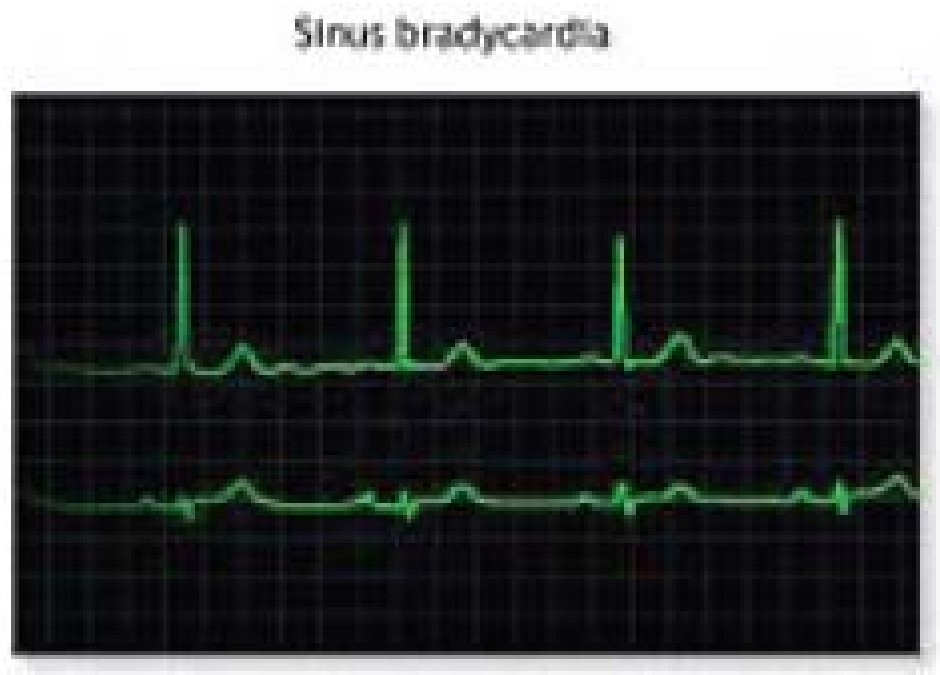

**Figure 10** Empirical data (ECG) showing slow heart rate or (sinus) Bradycardia. The plots are borrowed from [6] which is open access.

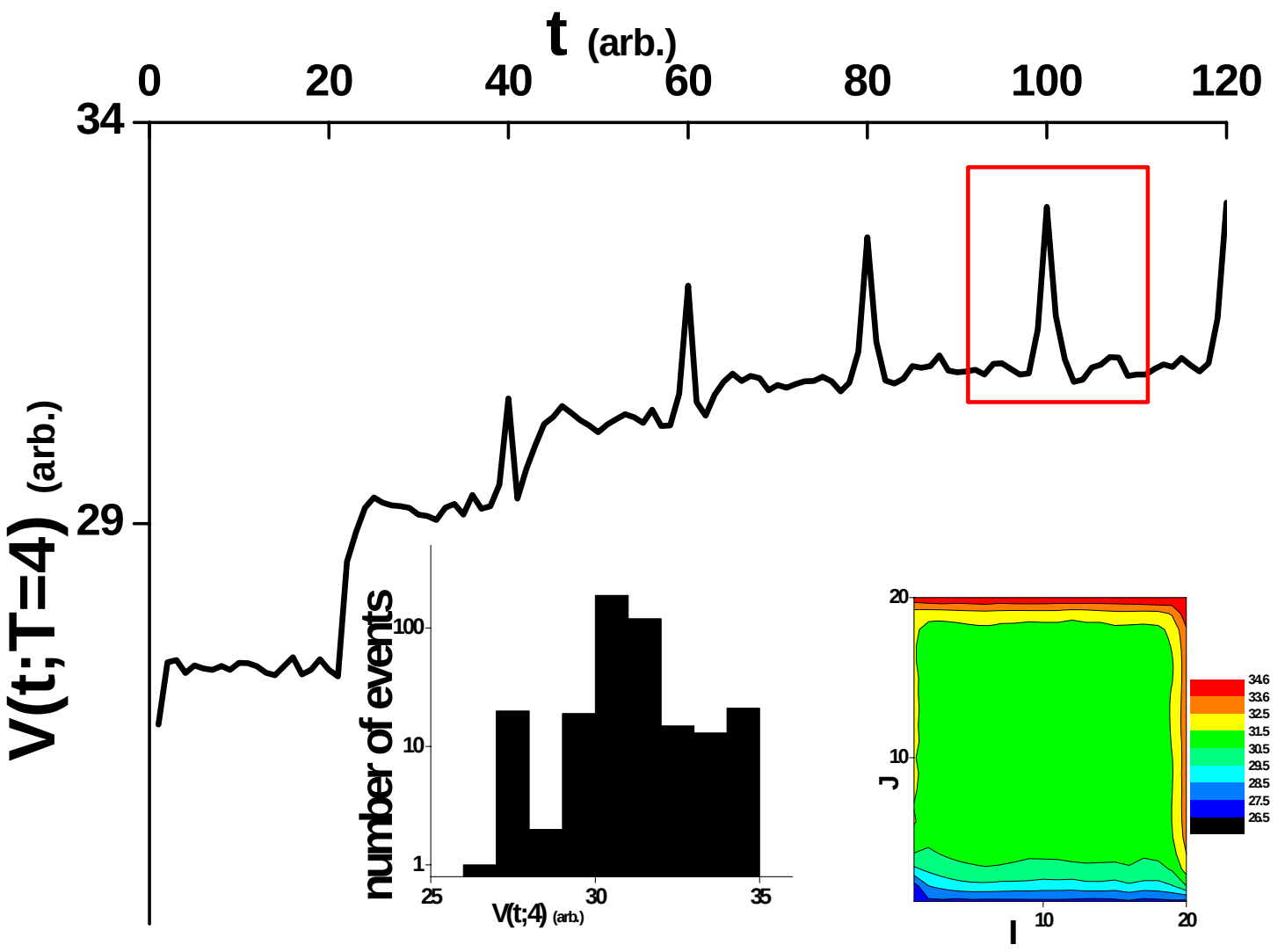

**Figure 11** Same as the Fig. 7 but for T=4 and with the feedings involving non random and random components (B=5 and R=0.2, respectively; Table I). The train of signals is (not computed but) simulated in terms of 4 secular equations (Eq. (1)).

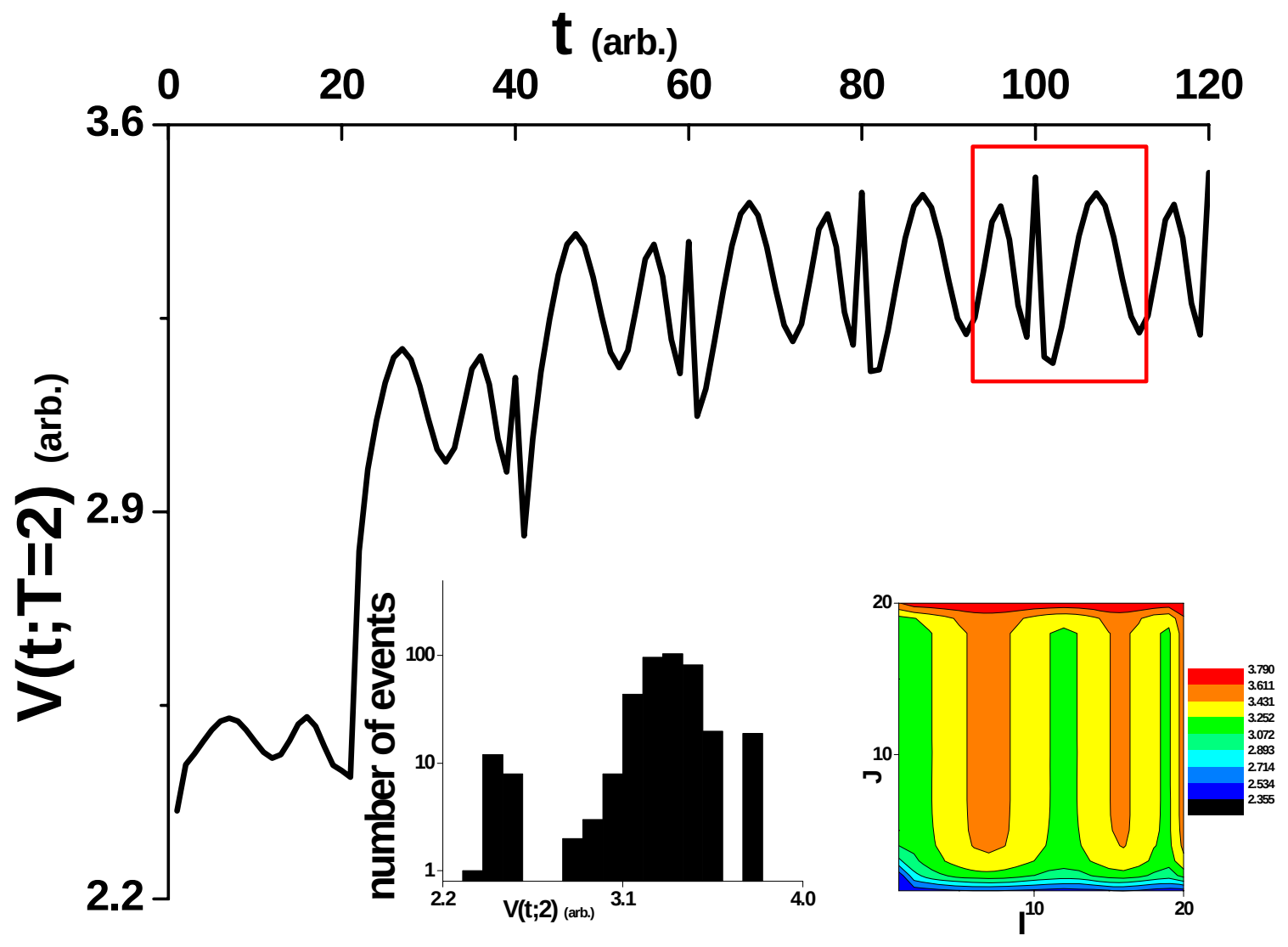

**Figure 12** Same as the Fig. 7 but for B=1 where no random parameter is used and the plot is computed in terms of 2 secular equations (Eq. (1)).

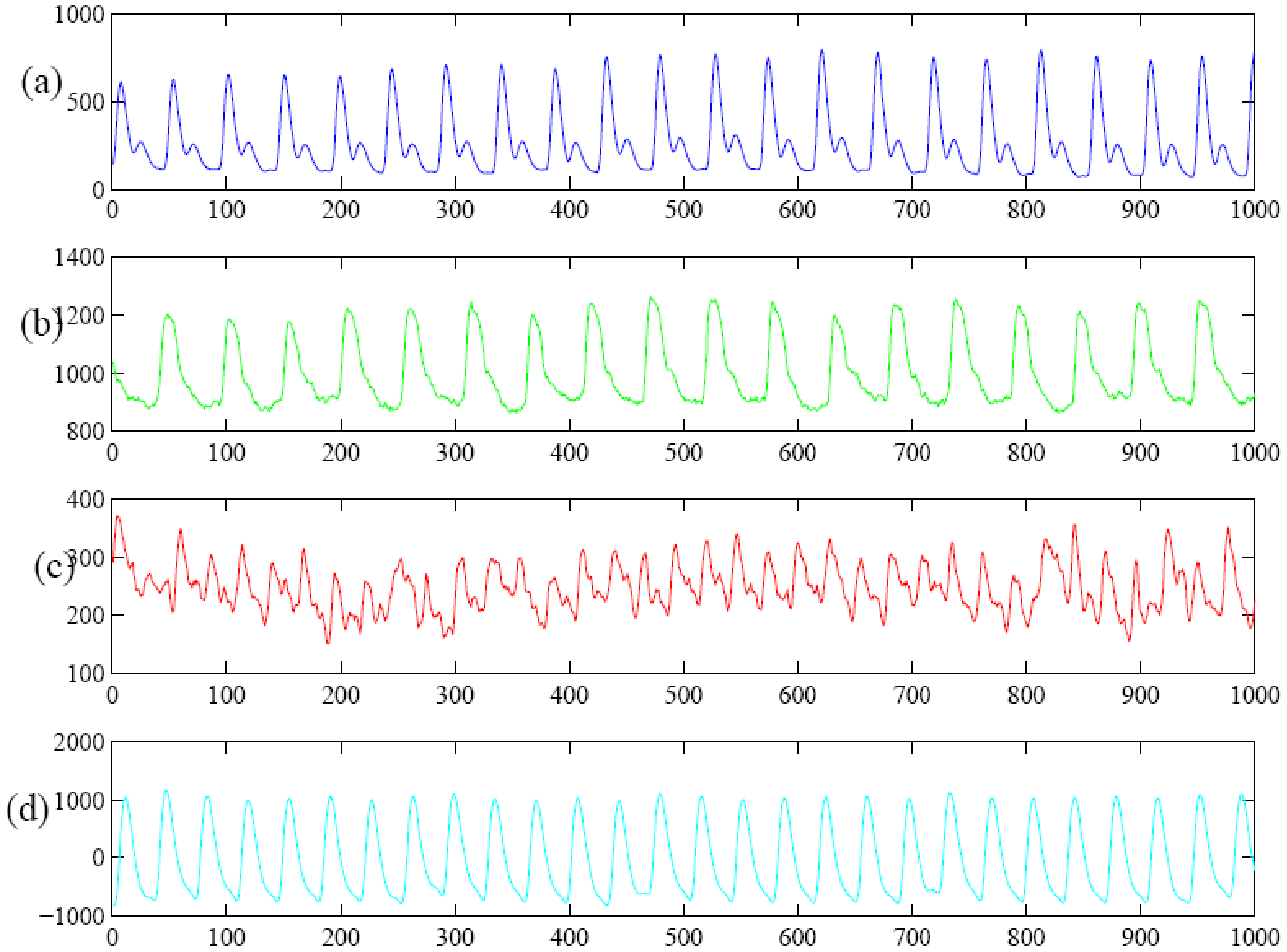

**Figure 13** The empirical data for human pulse waveform recorded with photo-plethysmography, which are borrowed from [2] open access. Four recordings of human pulse waveform are for the following physiological conditions: (a) normal, (b) quasi-stable, (c) unstable, and (d) post-operative (stable).

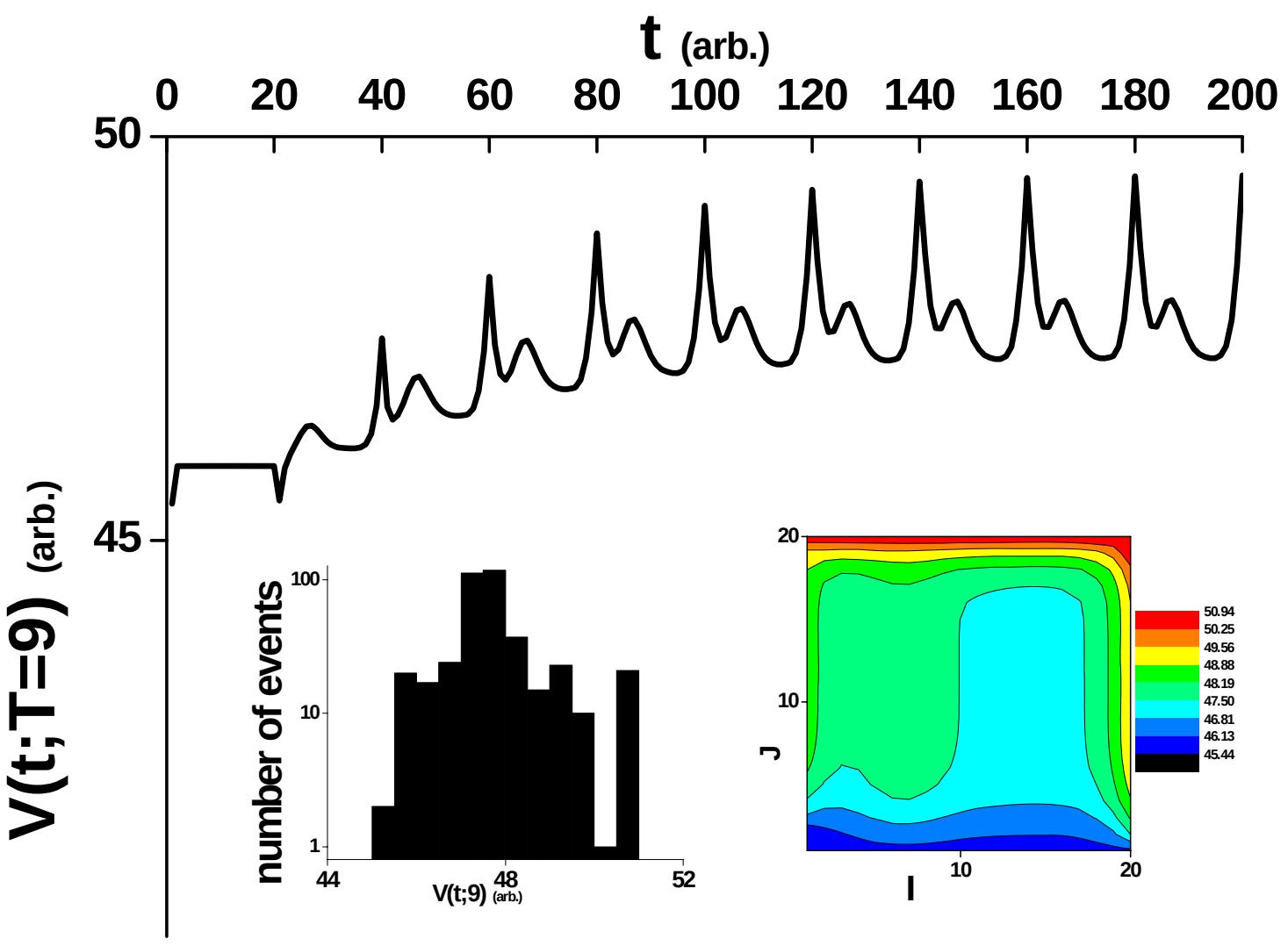

**Figure 14 (a)** Model heart signals for T=9, where the initial signal is composed of identical pulses defined in the Eq. (8). The model parameters are given in the Table I. The insets show the distributions of the number of the model voltage amplitudes V(t;9) over the voltages (the histogram at left, where the vertical axis is logarithmic) and the same amplitudes for C(I,J;9) over the entries (I,J) of the representing (20x20) lattice (at right), respectively. The plot is (not simulated but) computed in terms of 9 secular equations (Eq. (1)).

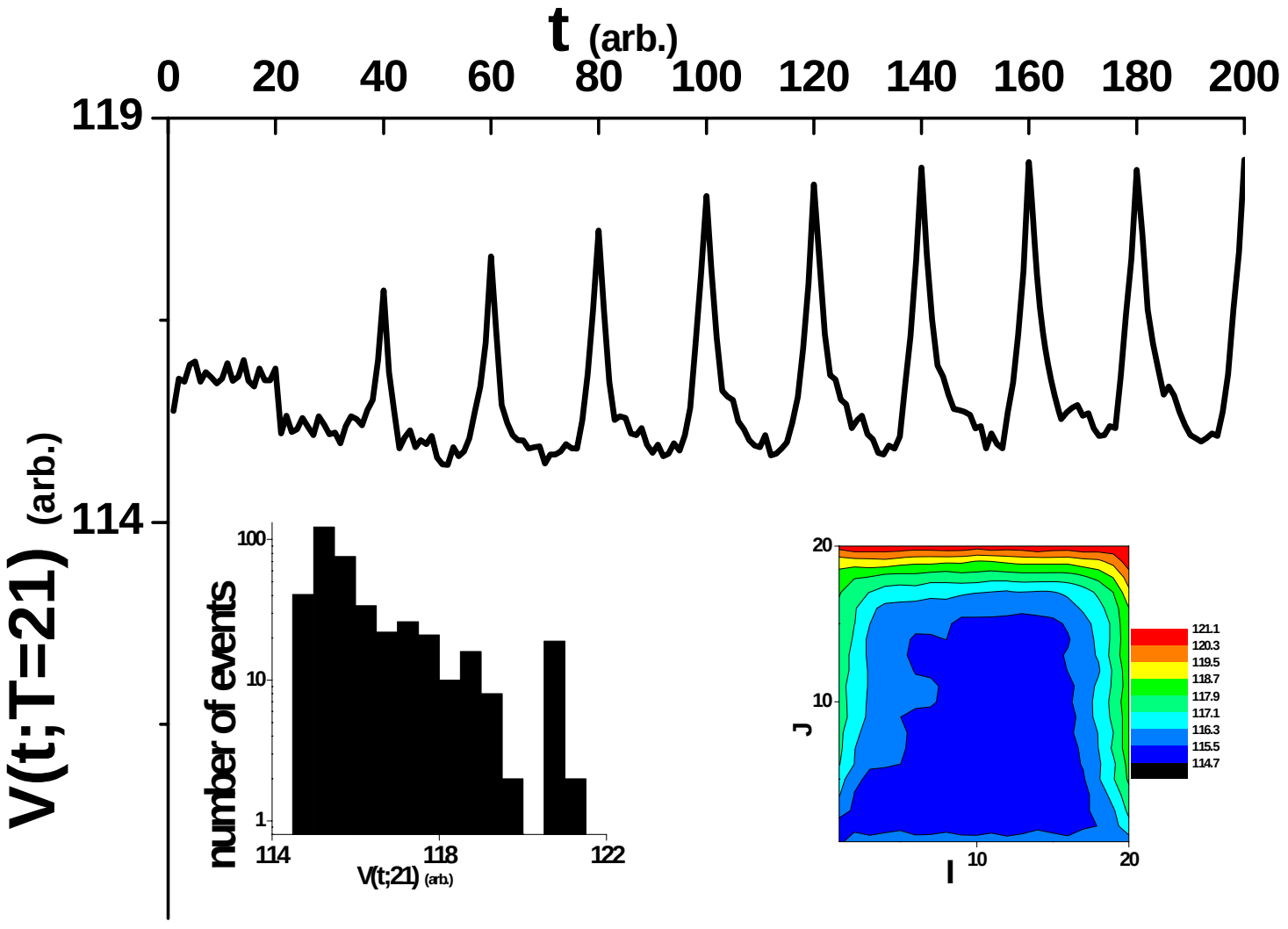

**Figure 14 (b)** Model signals as in the Fig. 14 (a) but for T=21, where some random parameter (R) is used for the feedings and the plots are simulated in terms of 21 secular equations (Eq. (1)).

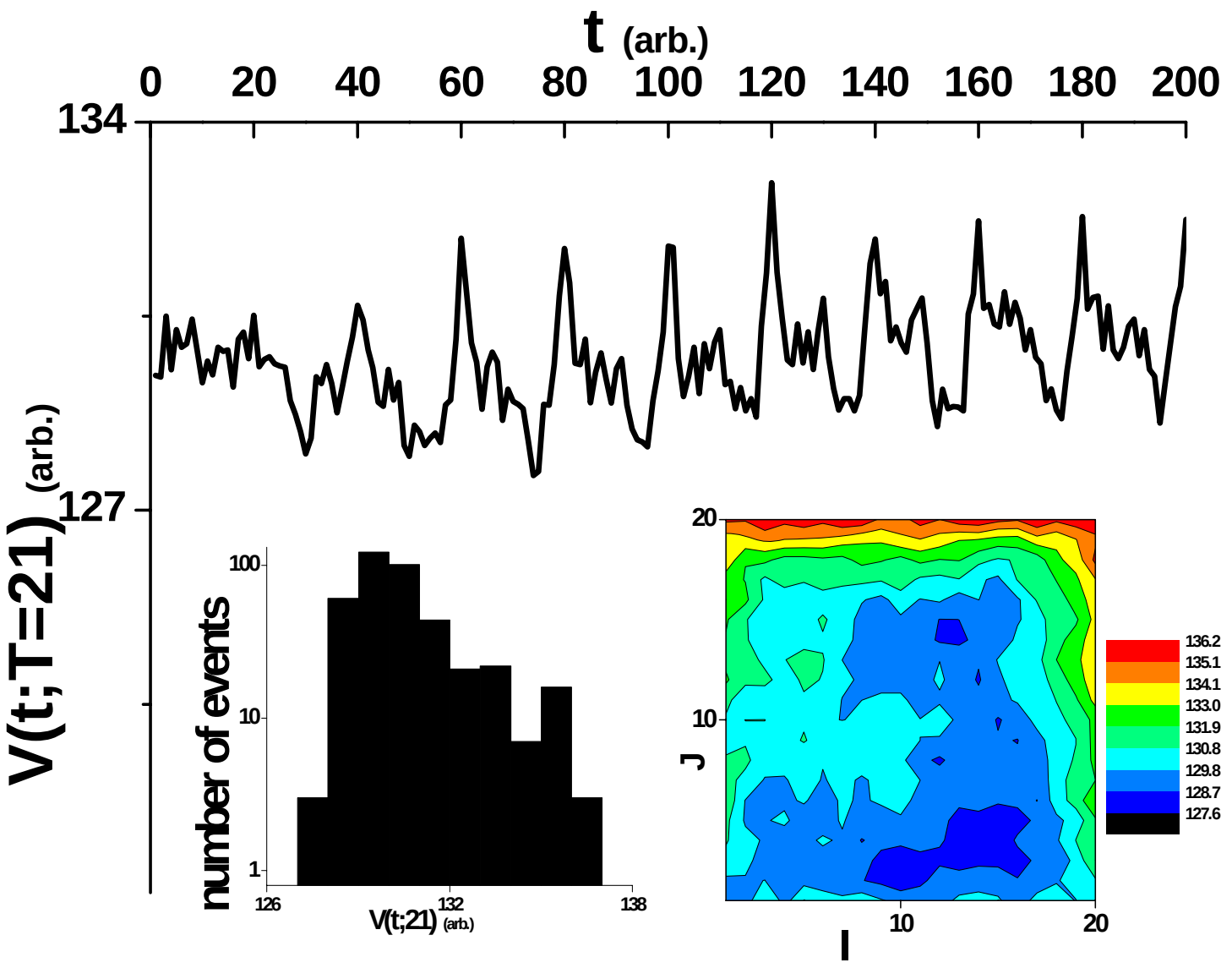

**Figure 14 (c)** Model signals as in the Fig. 14 (a) but for T=21, where a big random parameter (R) is used for the feedings and the plots are simulated in terms of 21 secular equations (Eq. (1)).

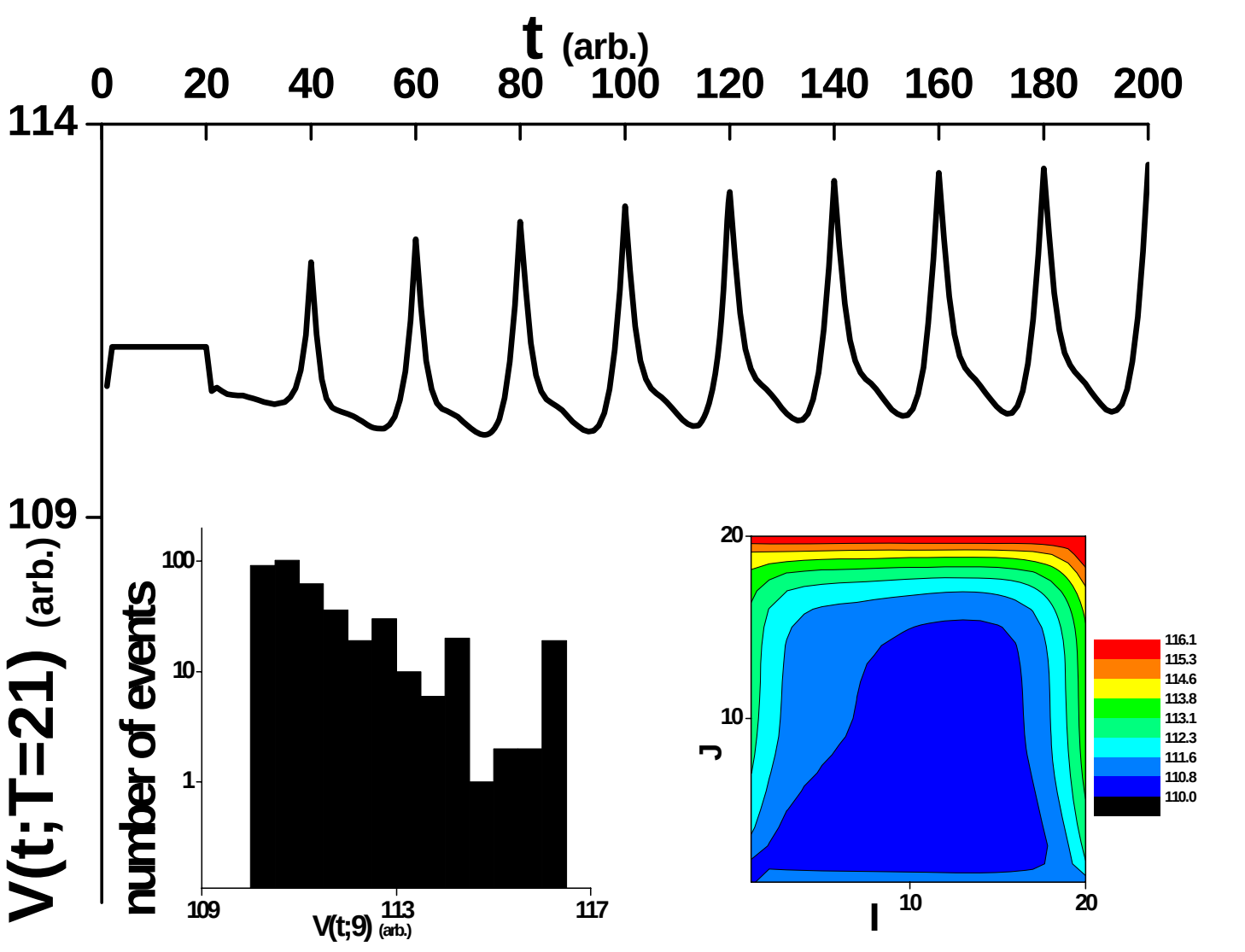

**Figure 14 (d)** Model signals as in the Fig. 14 (a) but for T=21. The plot is computed in terms of 21 secular equations (Eq. (1)), since no random parameter is involved here.

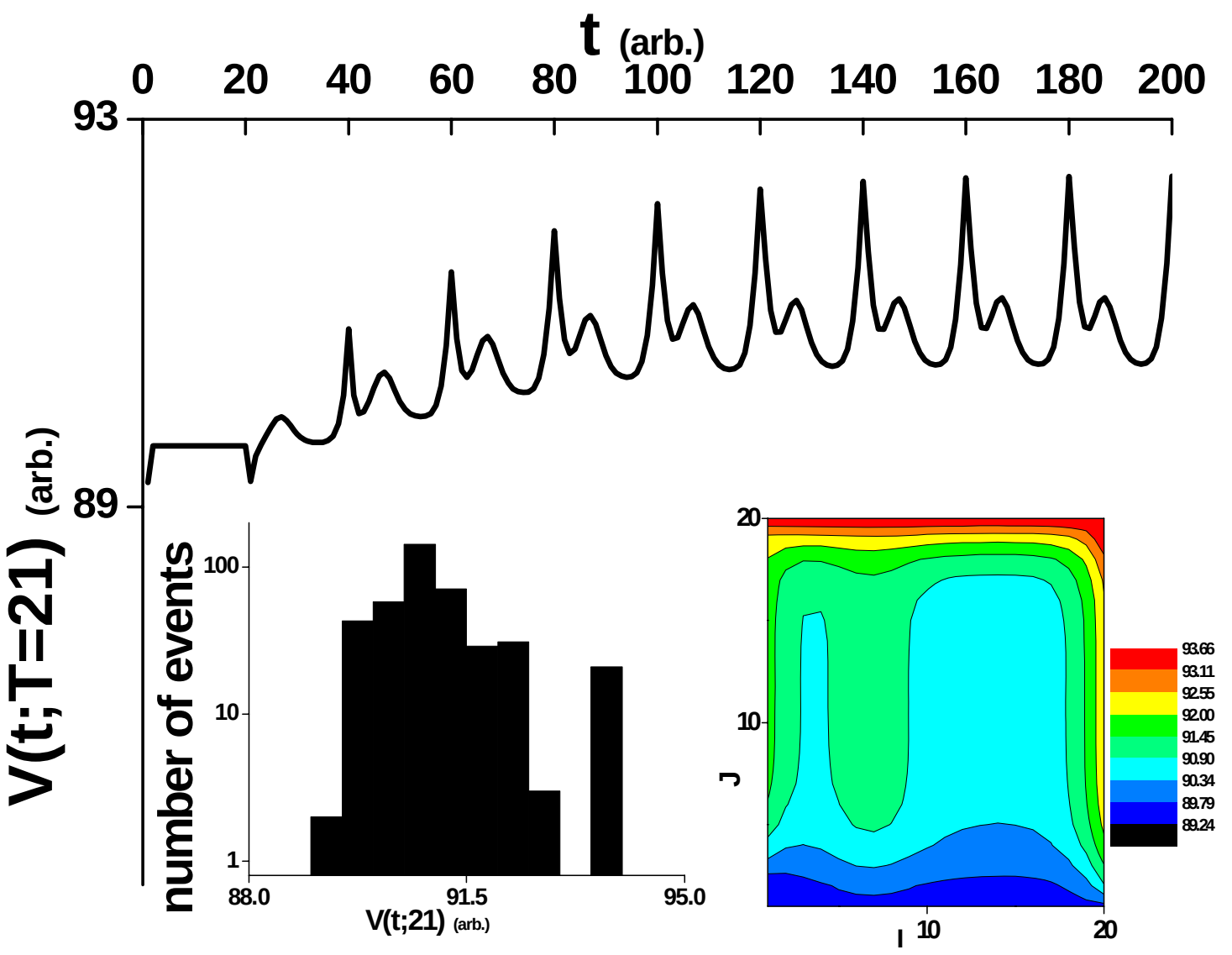

**Figure 15** Same as the Fig. 14 (a) but for T=21 and with smaller strength for the uniform connections (Table I). The plot is computed in terms of 21 secular equations (Eq. (1)).

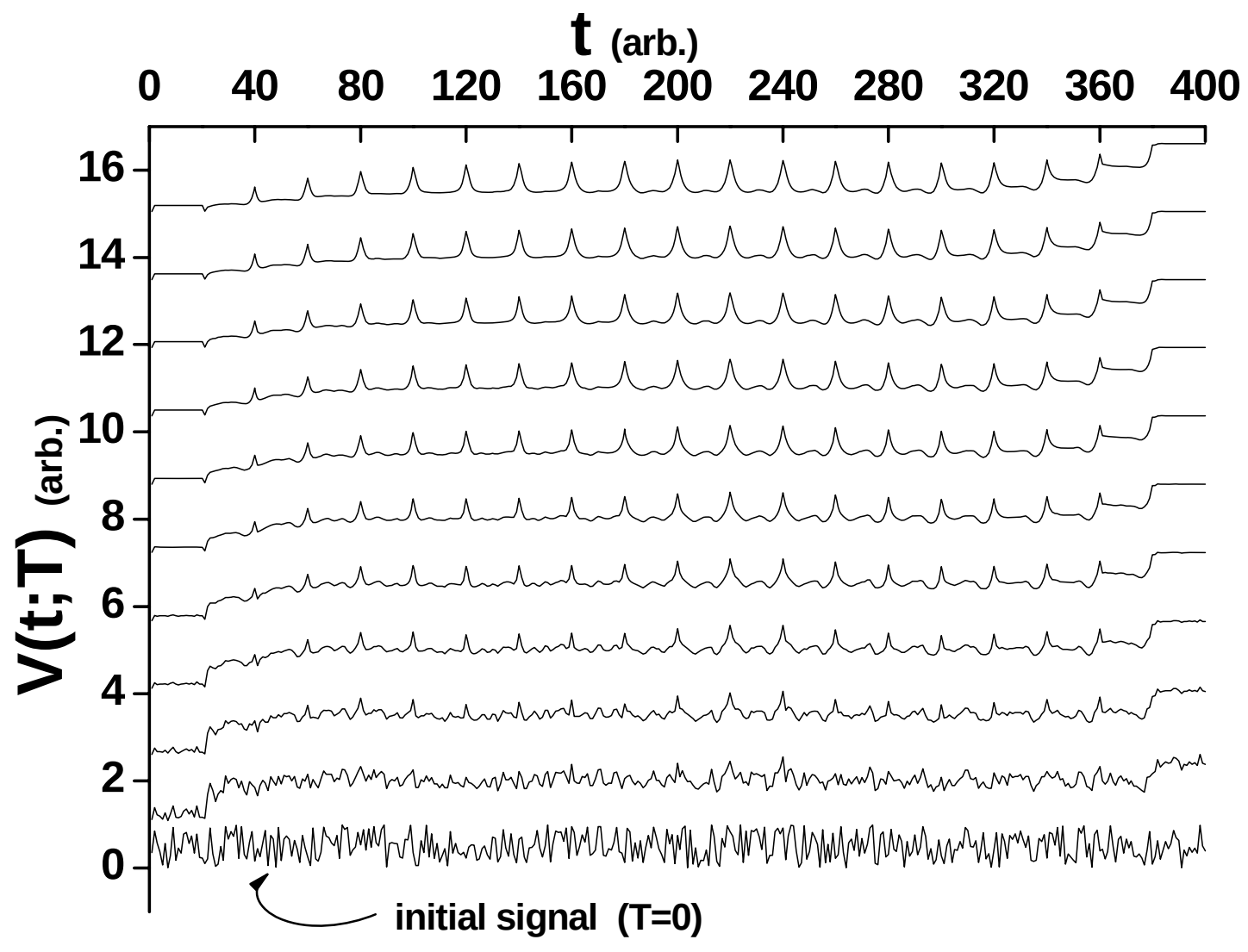

**Figure 16 (a)** Some model signals for various T which increases upward by one from zero for the initial signal (bottom) which is random (Eq. (9)) to 10 (top) where no random feeding is utilized (R=0;Table I).

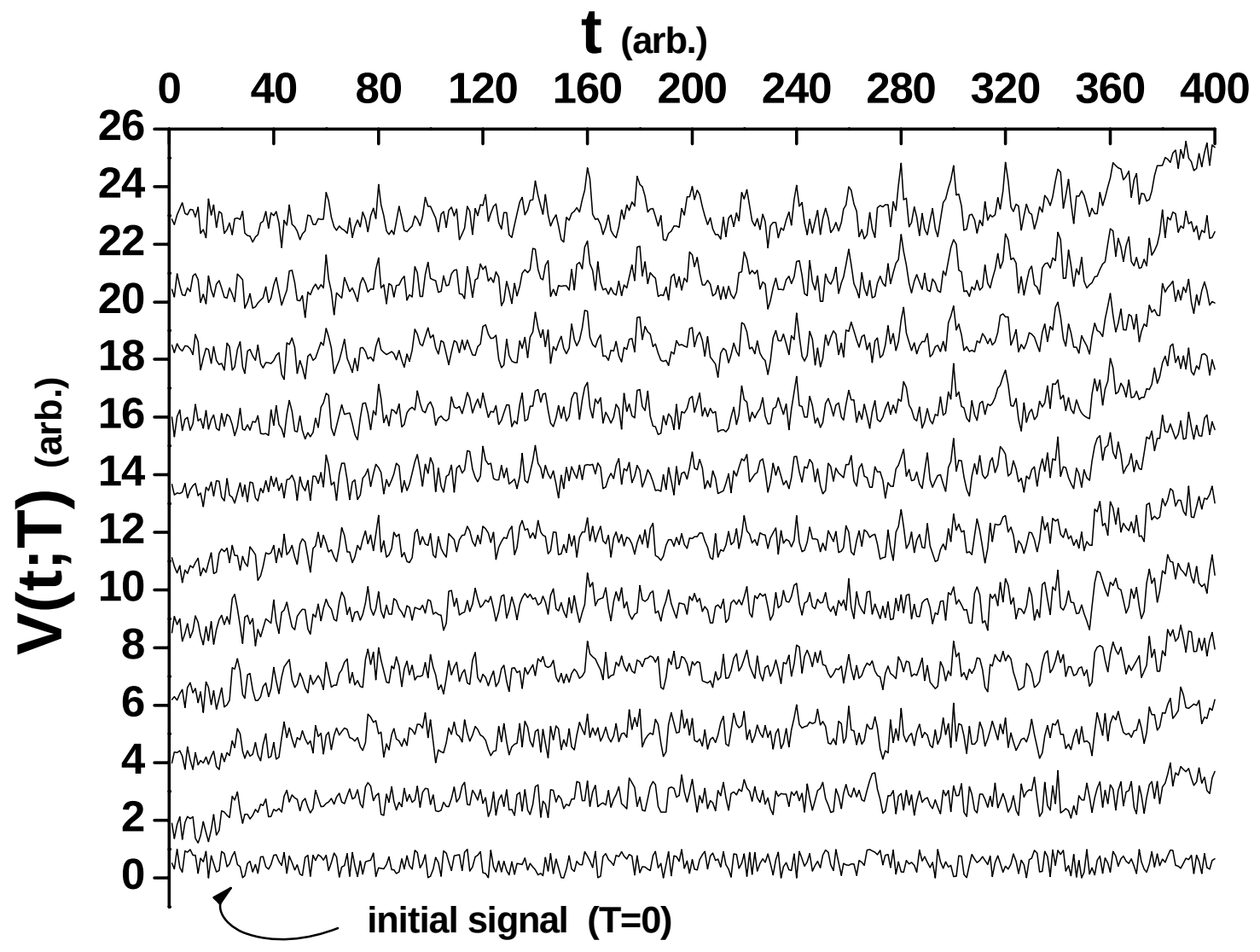

**Figure 16 (b)** Same as the Figure 16 (a) but the random and non random components of the feedings are equal (B=R=1; Table I) here.

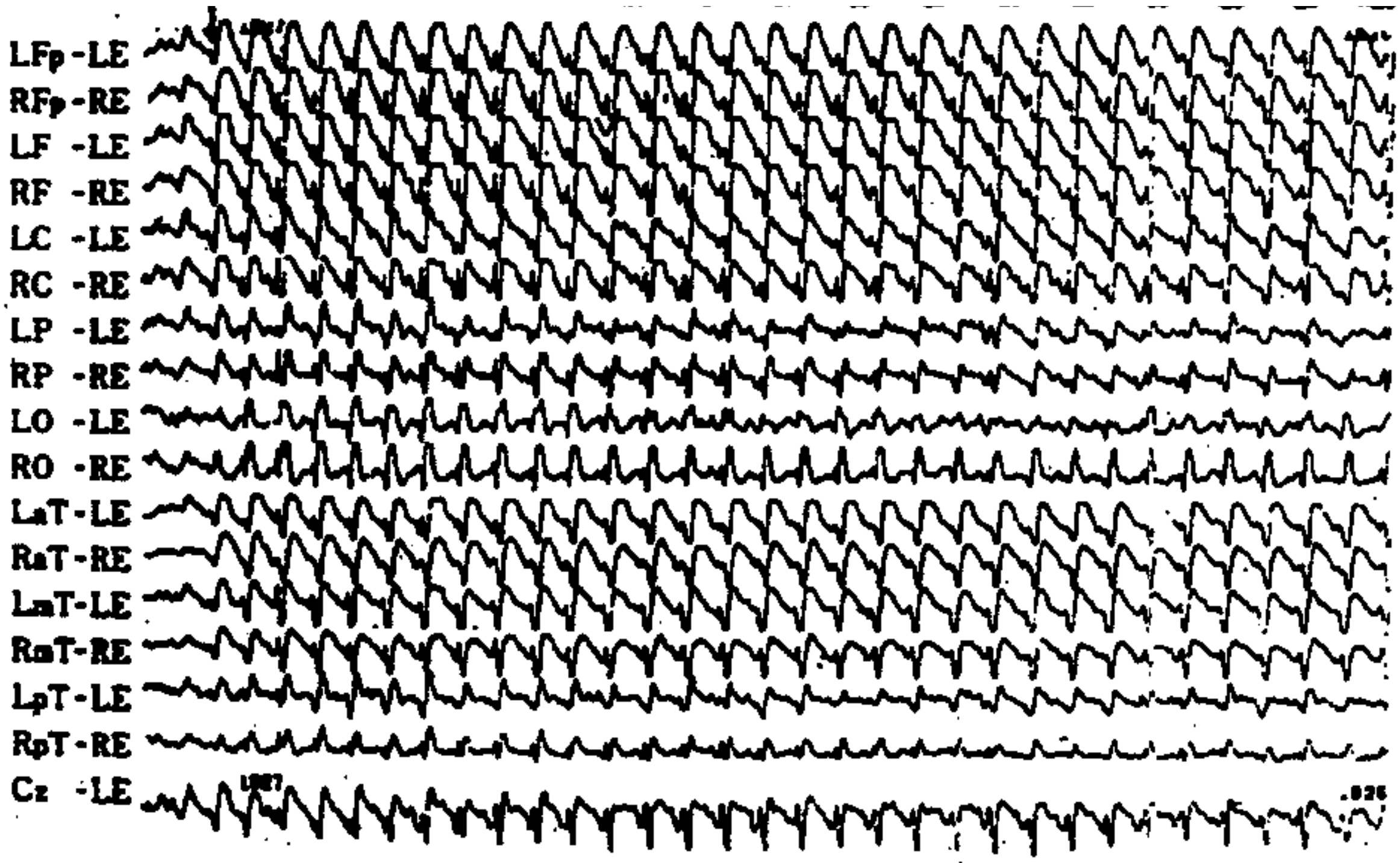

**Figure 17** Various EEG signals indicating absence seizure.

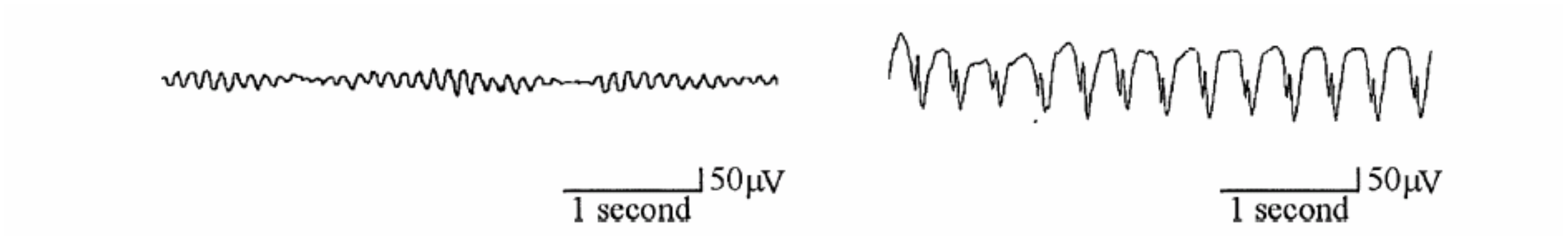

**Figure 18** EEG signals about two epochs of absence seizure.

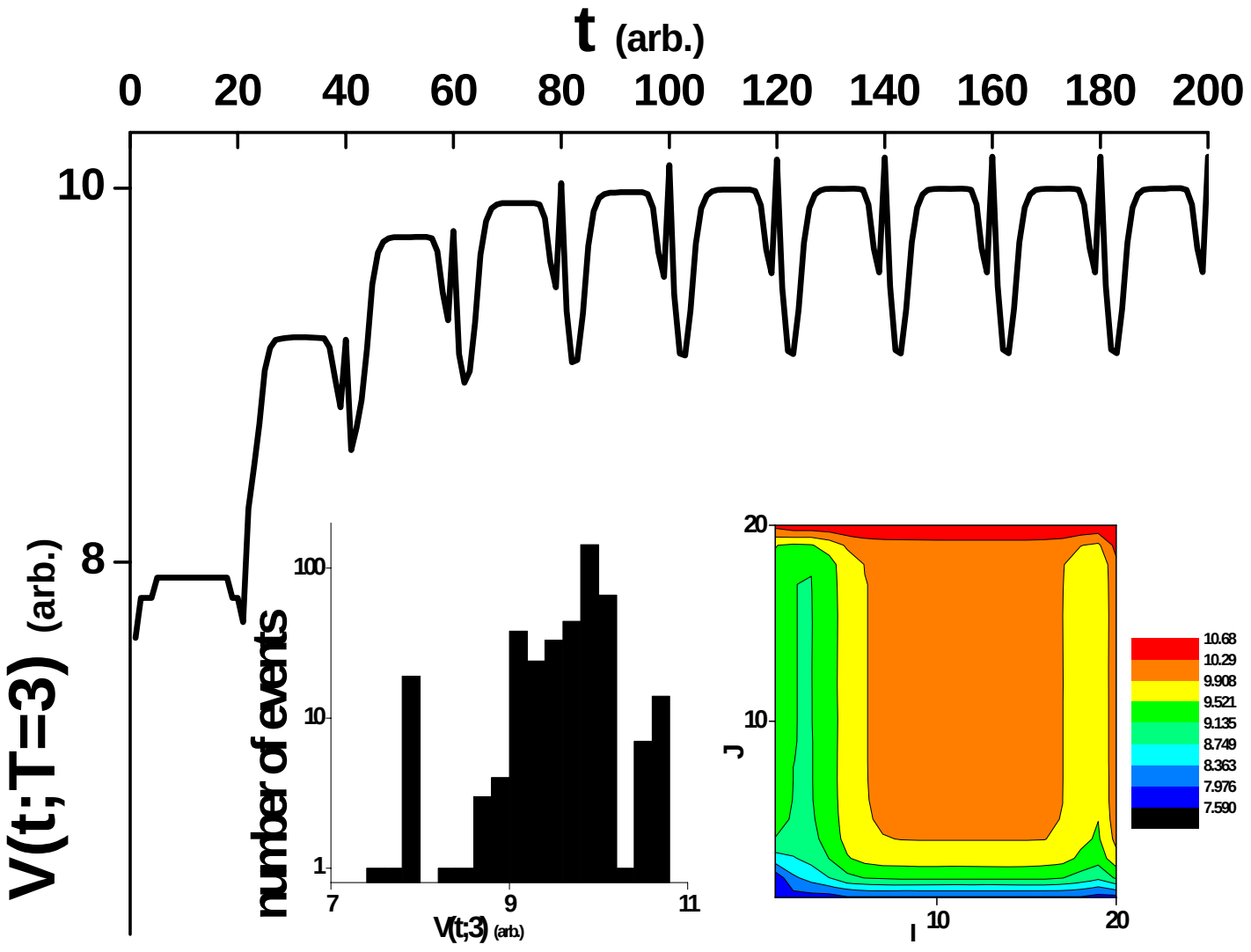

**Figure 19** Same as the Fig. 14 (a) but for T=3 and with a rectangular (step or well) periodic initial signal (Eq. (10)). The parameters are given in the Table I.

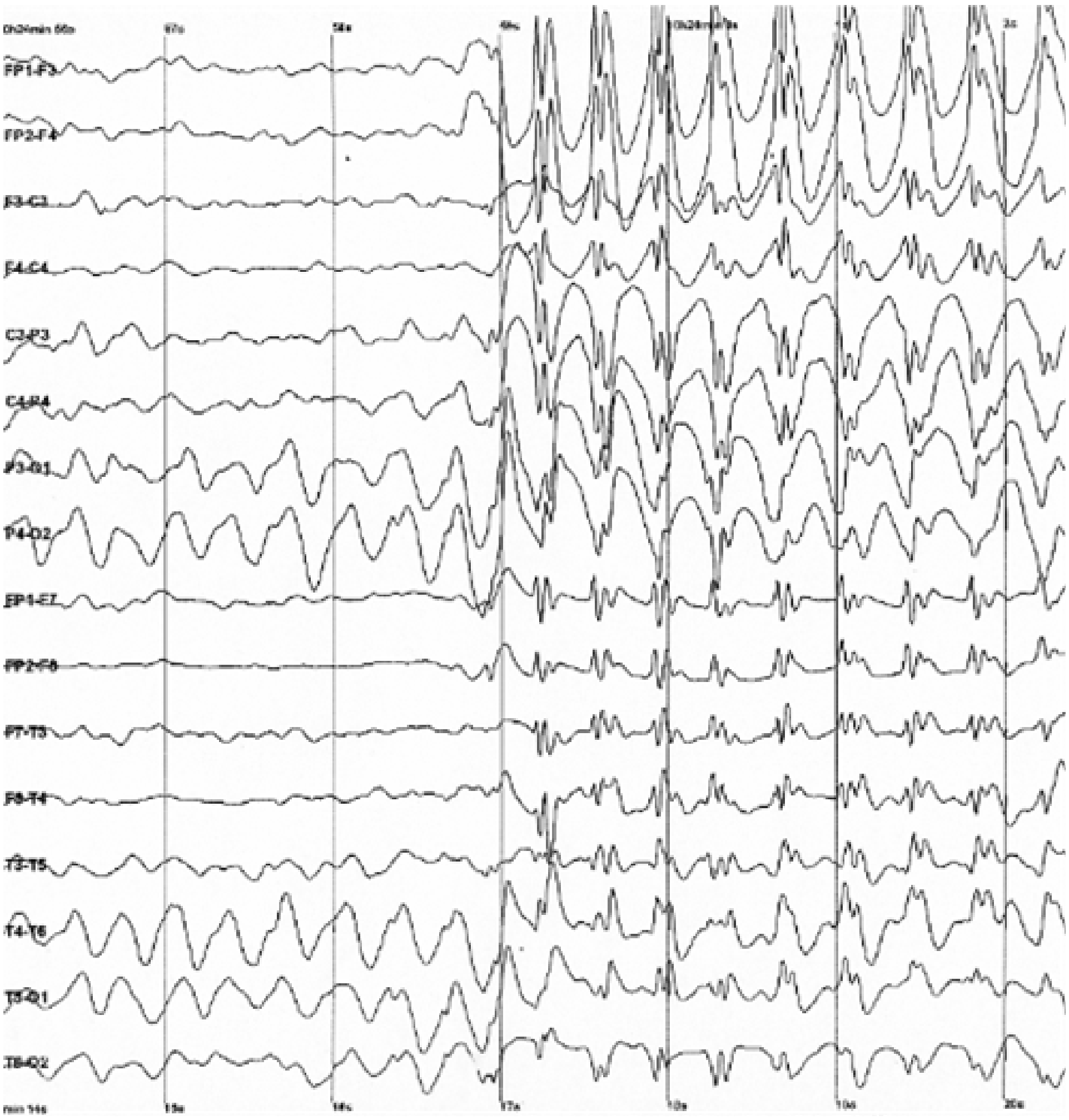

**Figure 20** Generalized 3 Hz spike and wave discharges in a child with childhood absence epilepsy which is borrowed from http://en.wikipedia.org/wiki/Image:Spike-waves.png (open accees).

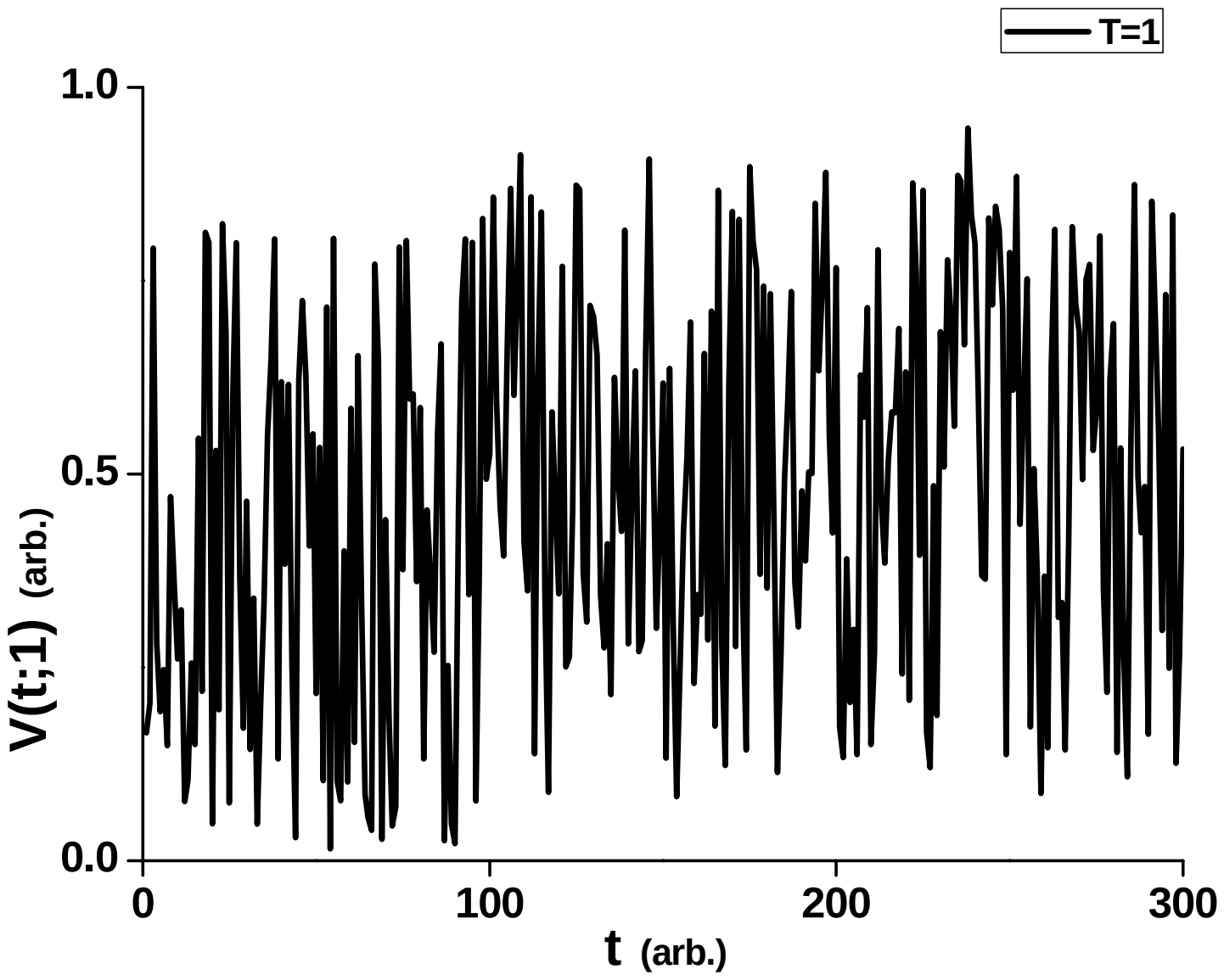

(a)

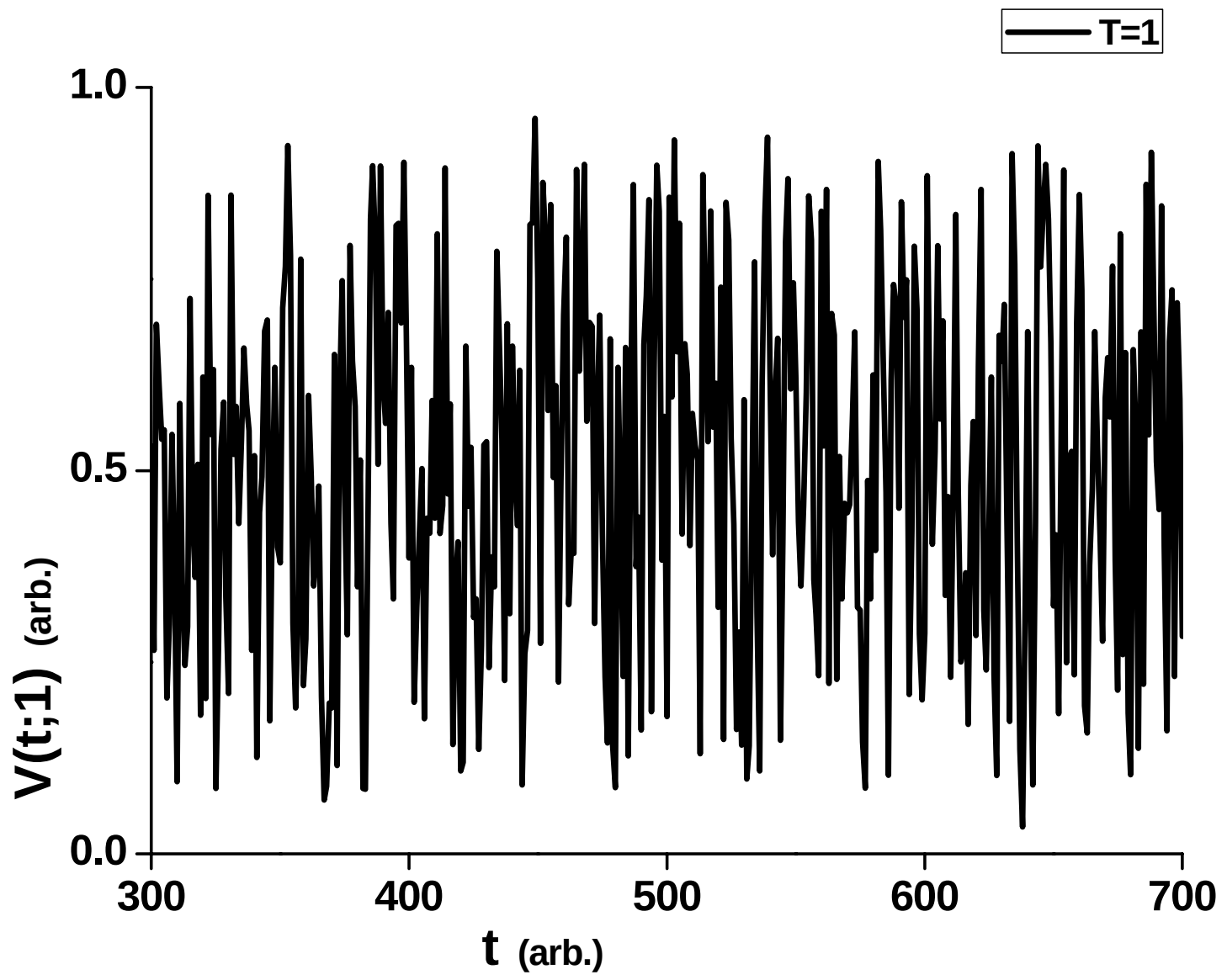

(b)

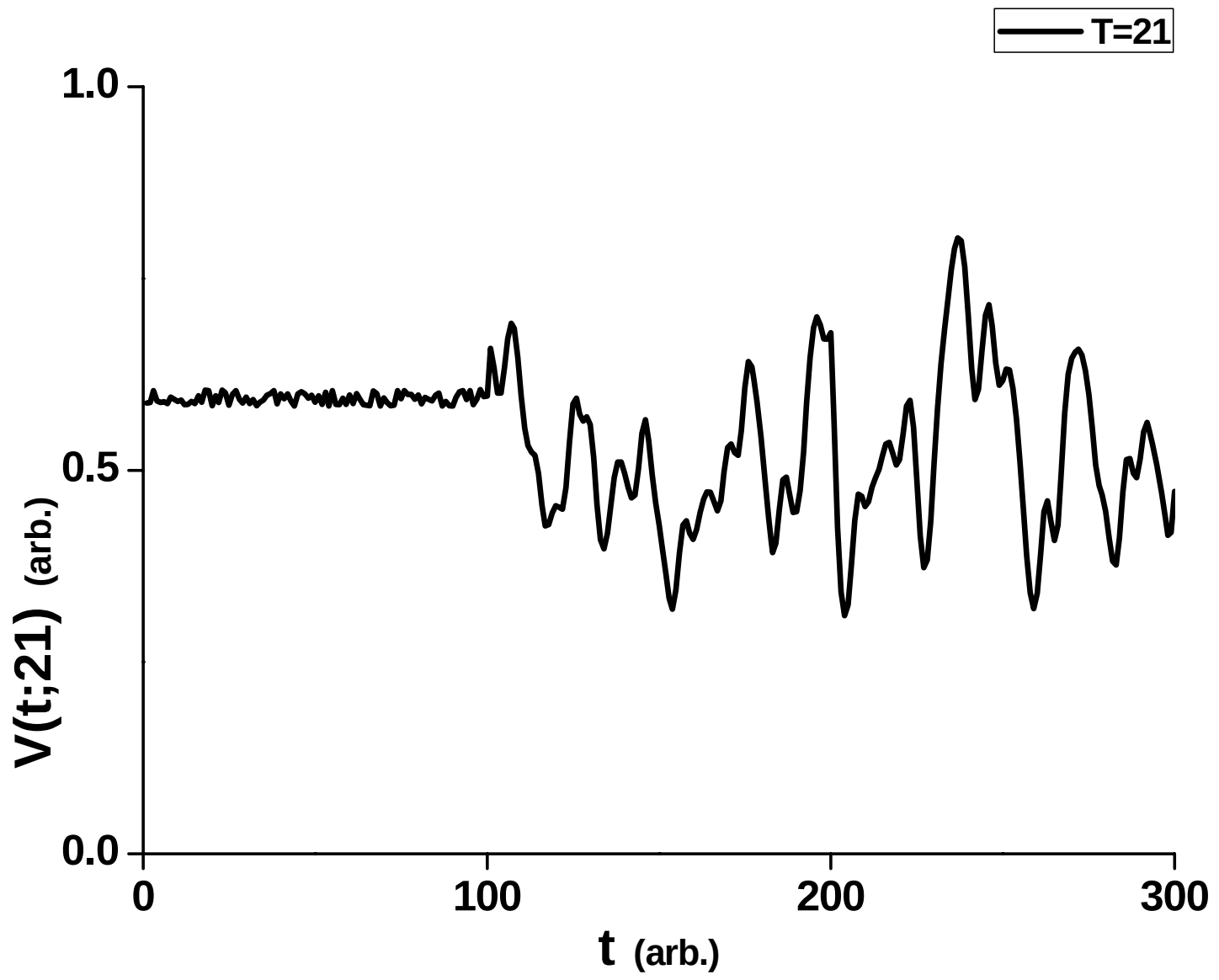

(c)

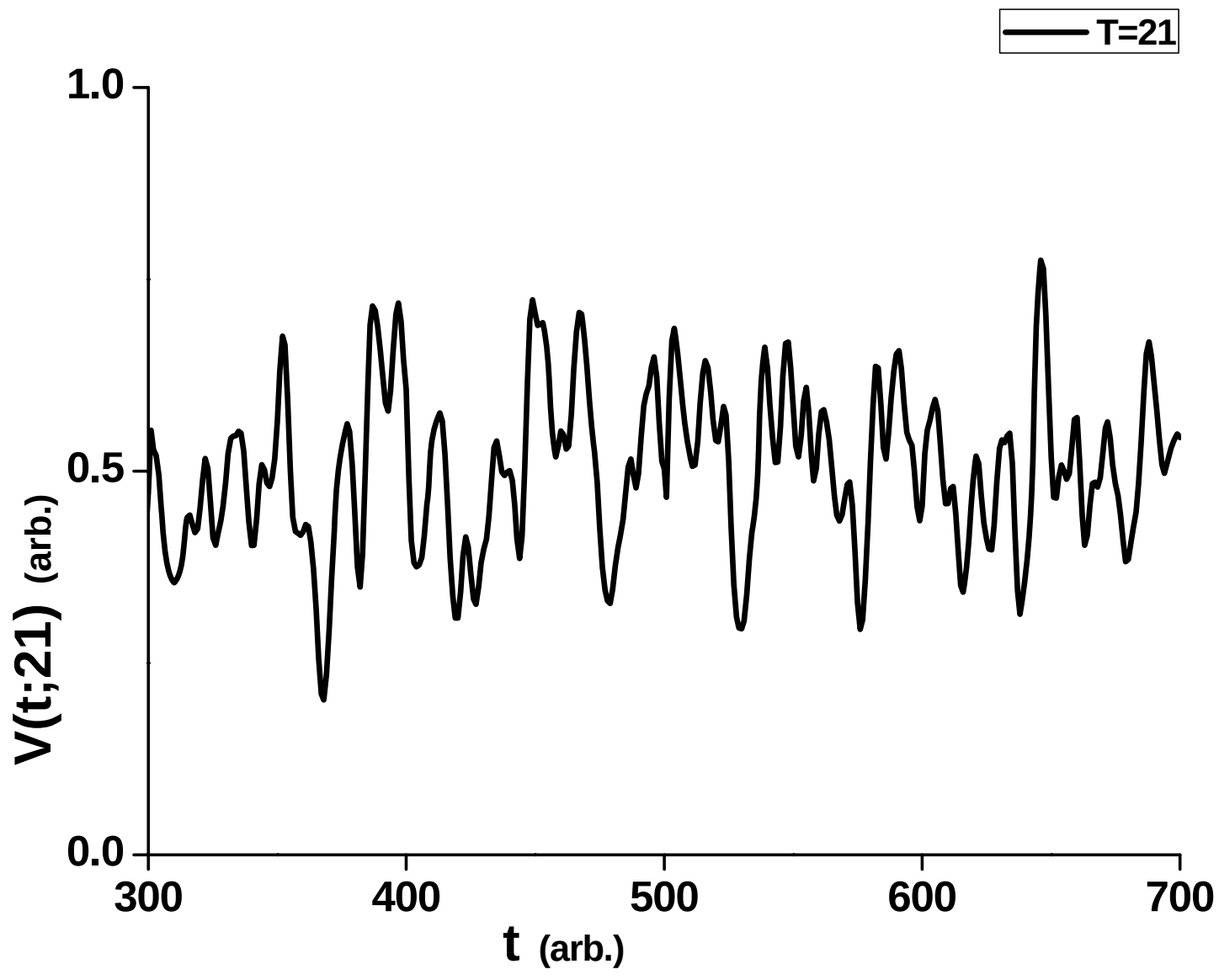

(d)

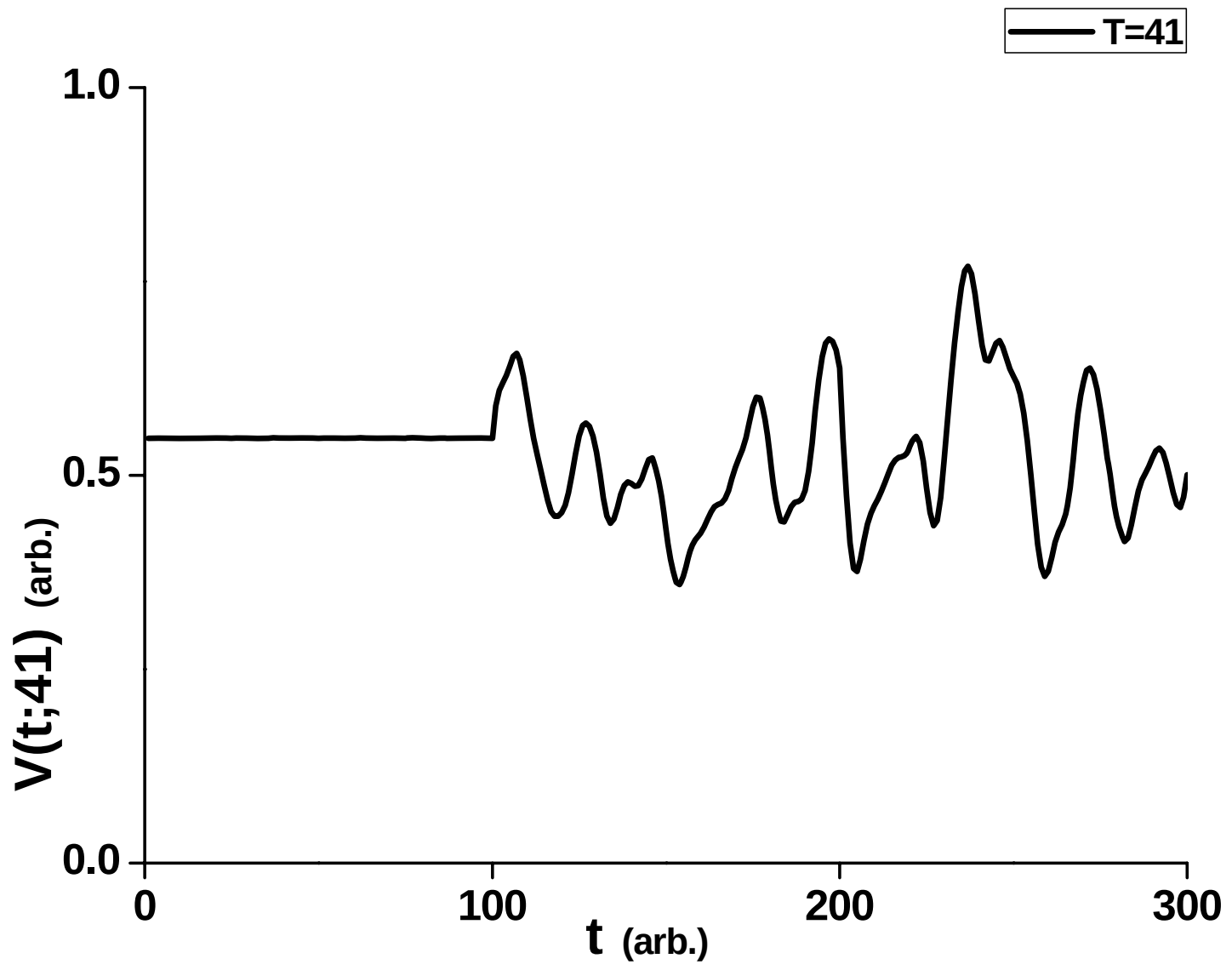

(e)

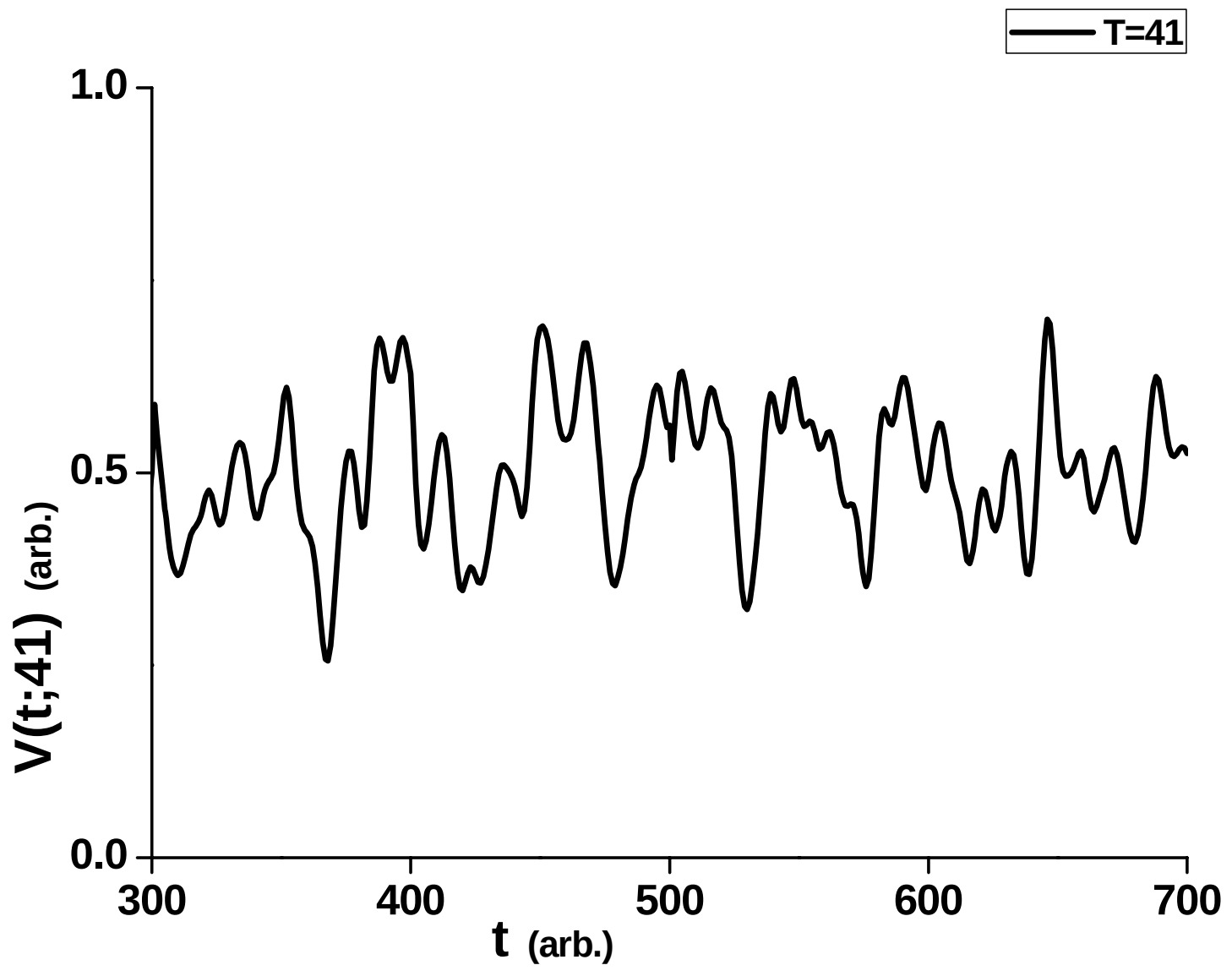

(f)

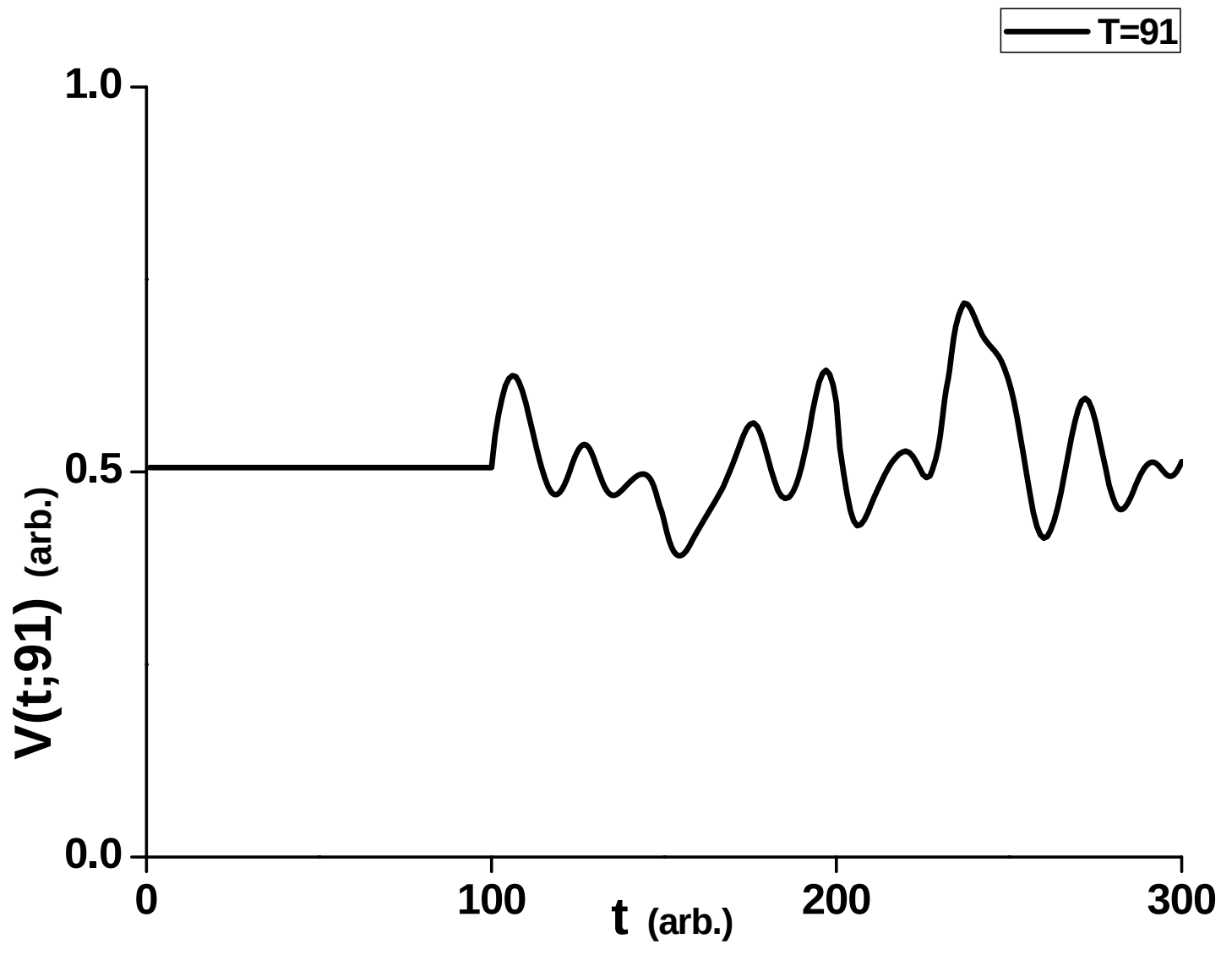

(g)

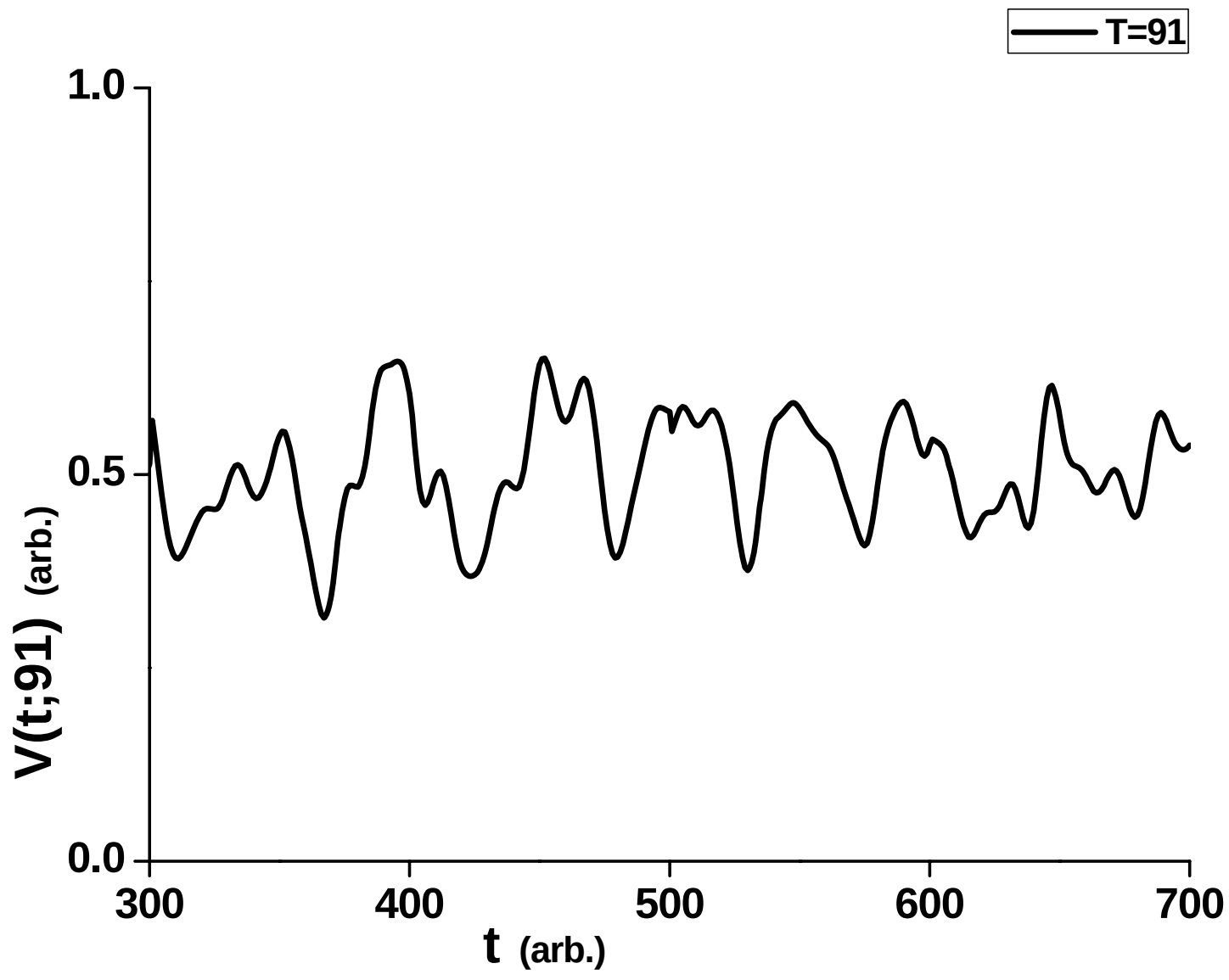

(h)

**Figure 21** Some model signals where random initial waves (Eq. (9)) and non uniform connectivity (h=0, v=0.1) are used with N=100; (a) and (b) are for T=0; (c) and (d) are for T=21; (e) and (f) are for T=41 and (g) and (h) are for T=91 where different epochs for 0≤t≤300 and 300≤t≤700 are shown, all respectively.

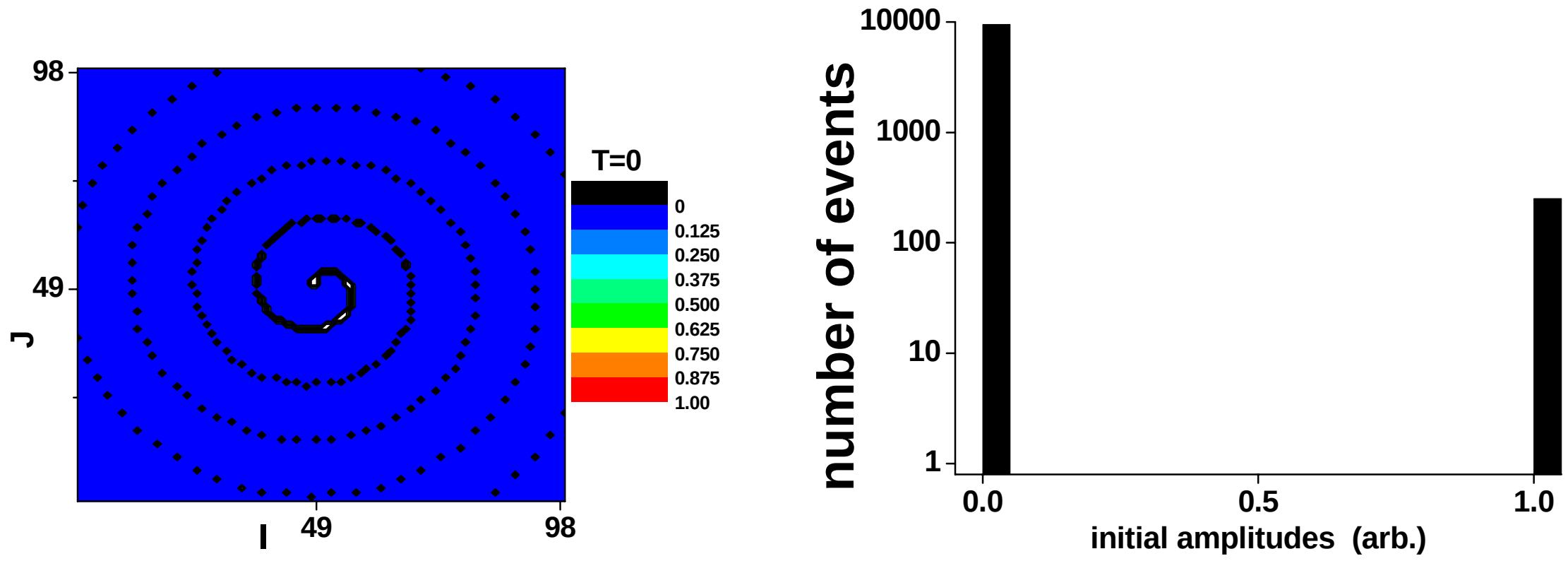

(a)

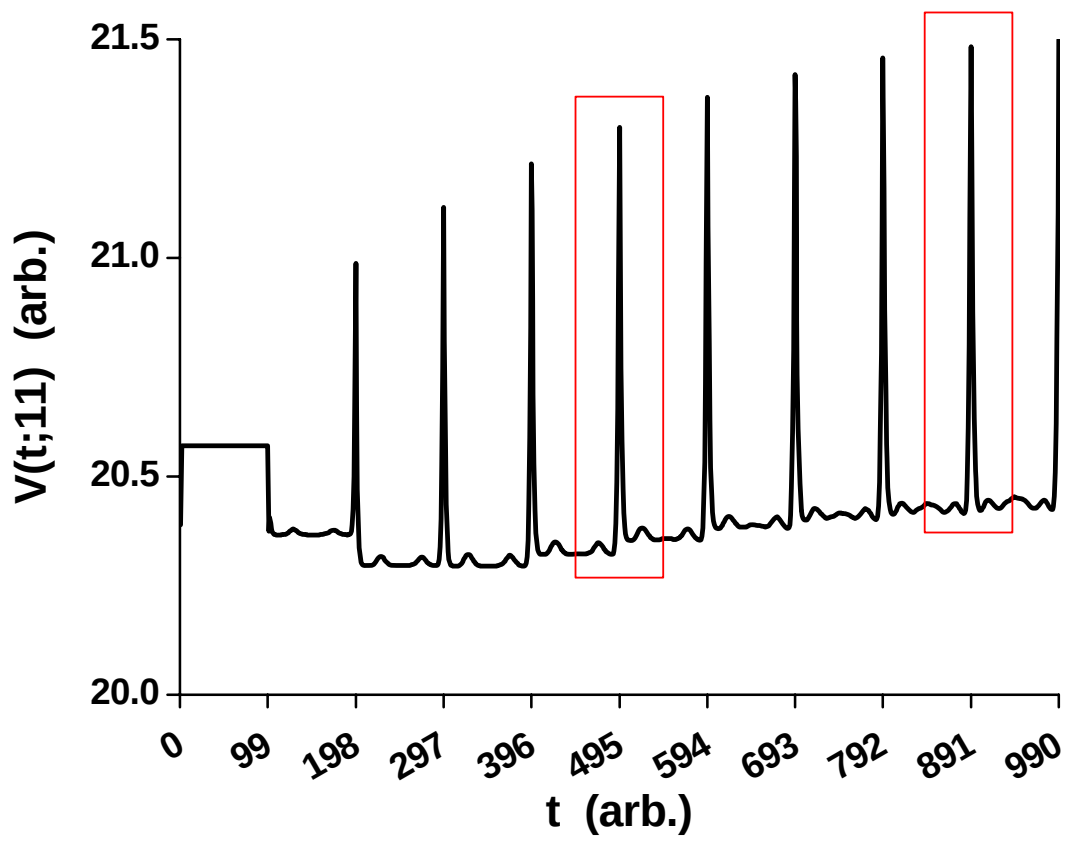

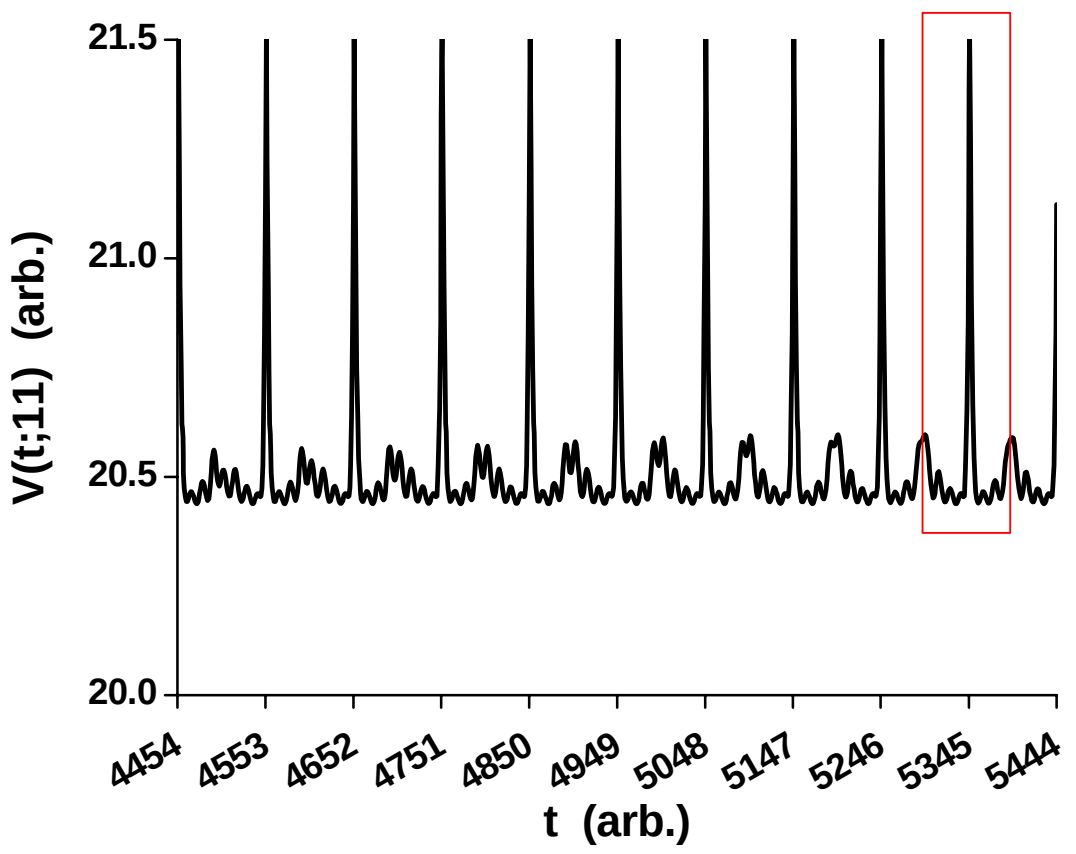

(**b**)

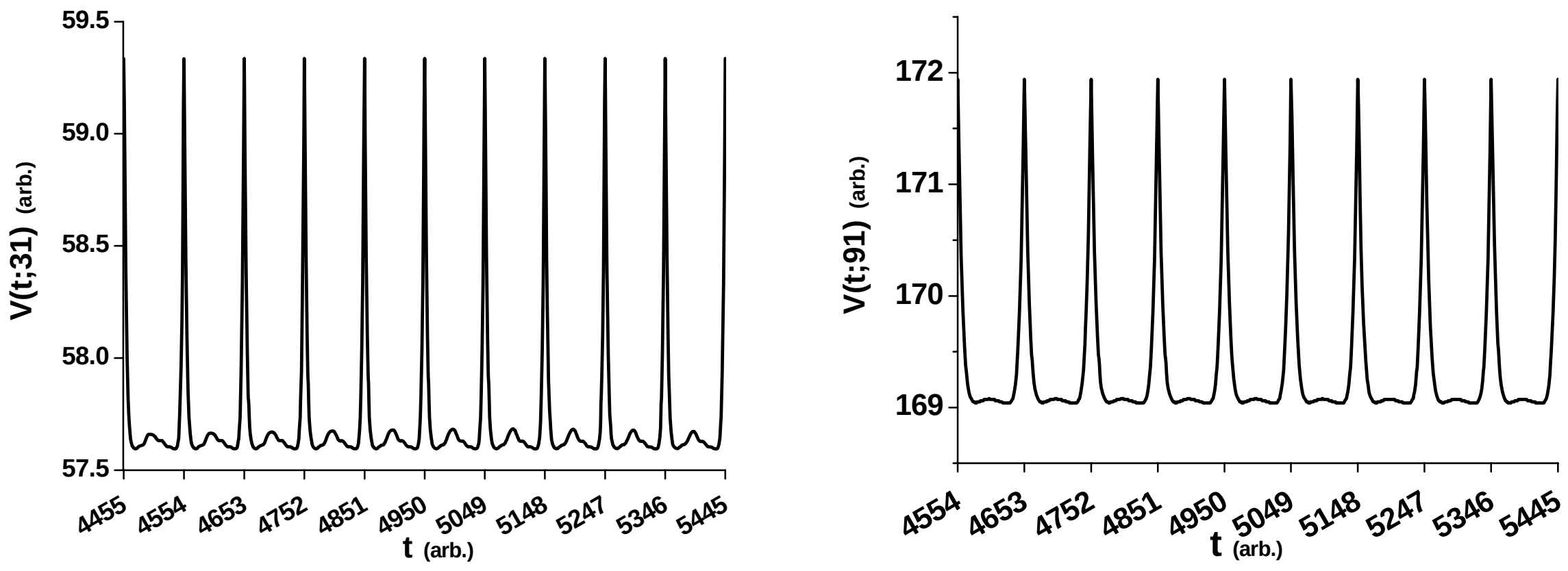

(**c**)

**Figure 22 (a) – (c)** Some model signals with an initial spiral wave. Please see the Eq. 17 and the text in the paragraph next to the mentioned equation for the meaning of the model parameters. (**a**) The initial configuration of the representing lattice in terms of a spiral (left) wave and the histogram for the distribution of the amplitudes in C(I,J;0) or the voltages V(t,0) (right) where the number of the entries with non zero amplitudes is 252 (right, histogram). (**b**) V(t,11) within different epochs where the variation in the topography (design, similarity) of the beats is clear which is called spatiotemporal change (or variation) in the related experimental literature. All of the frames in color online (red) have 99 units of time for t. (**c**) Evolution of the model signals for T=31 (left) and T=91 within the same epoch; 4455≤t≤5445.

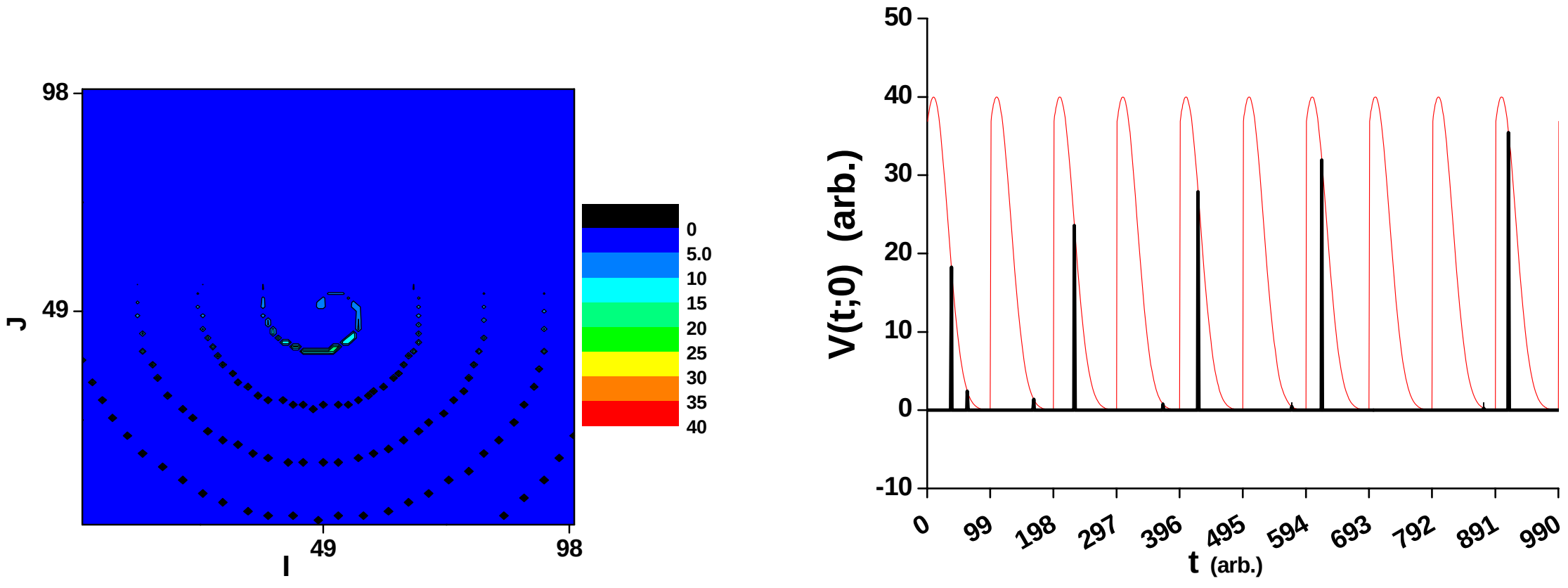

(**a**)

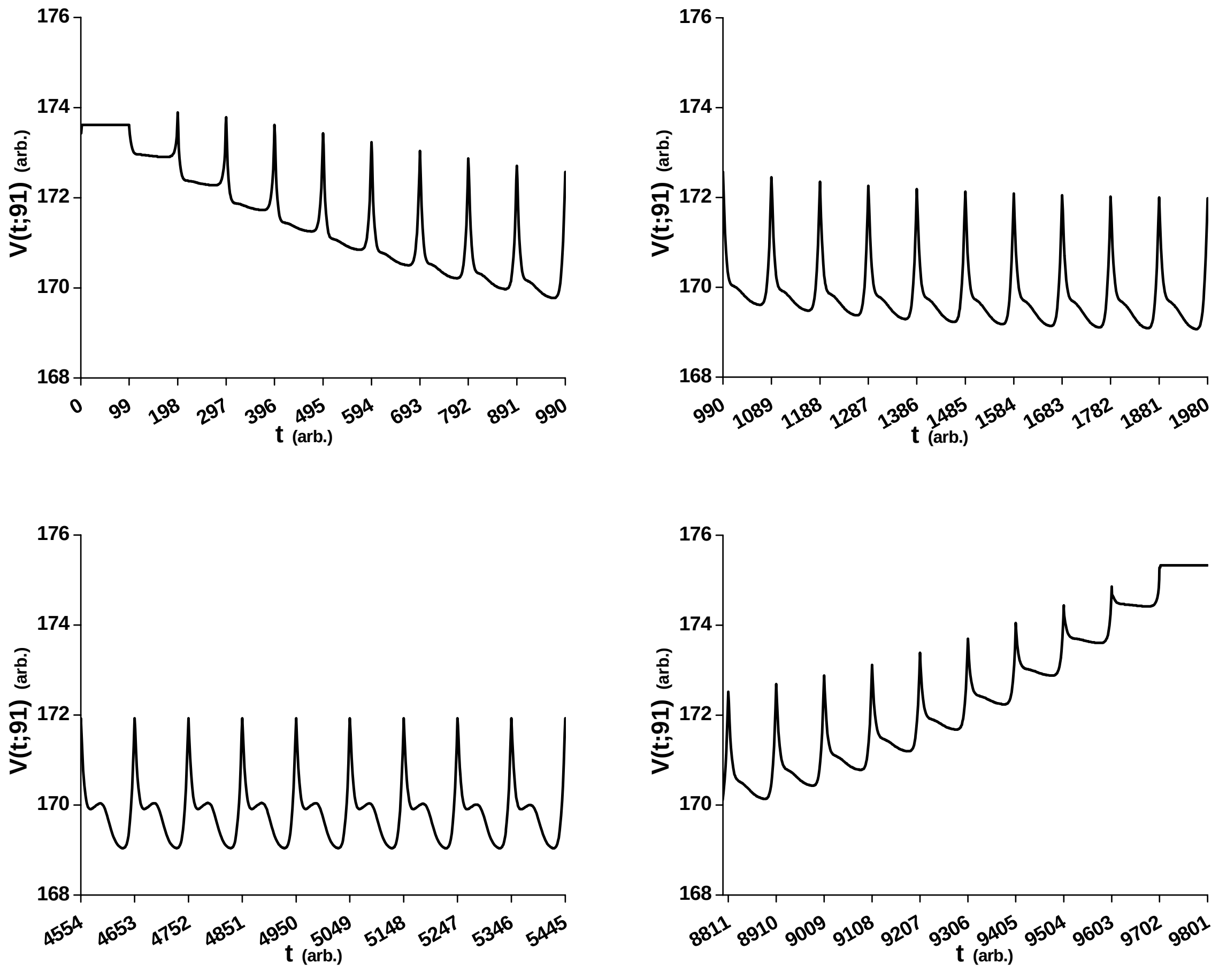

(**b**)

**Figure 23 (a) – (b)** Some model signals with initial spiral waves where the amplitudes are multiplied by a Gauss factor; please see the Eq. 18 and the text in the paragraph next to the mentioned equation for the meaning of the model parameters. (**a**) The initial configuration of the representing lattice in terms of the mentioned spiral wave (left) and the initial signal V(t,0) (right) where the envelope (color online) is Gauss and thus, the initial signal is not uniform in the amplitudes. Please note that the number of the entries with non zero amplitudes is same as before (=252). (**b**) V(t,91) within different epochs; (up and left) $1 \leq t \leq 990$, (up and right) $990 \leq t \leq 1980$, (down and left) $4455 \leq t \leq 5445$ and (down and right) $4455 \leq t \leq 5445$ where the spatiotemporal changes are clear.

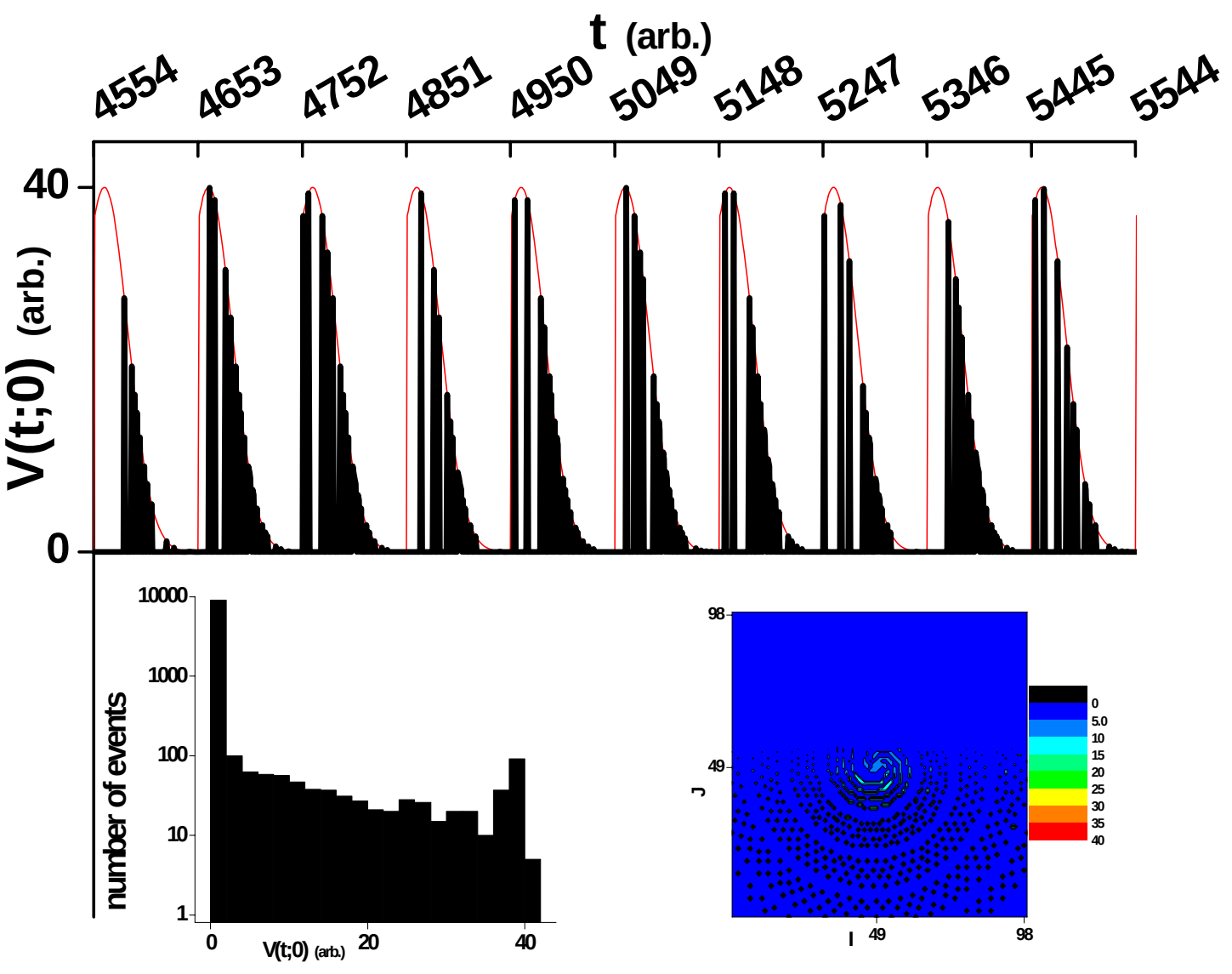

(a)

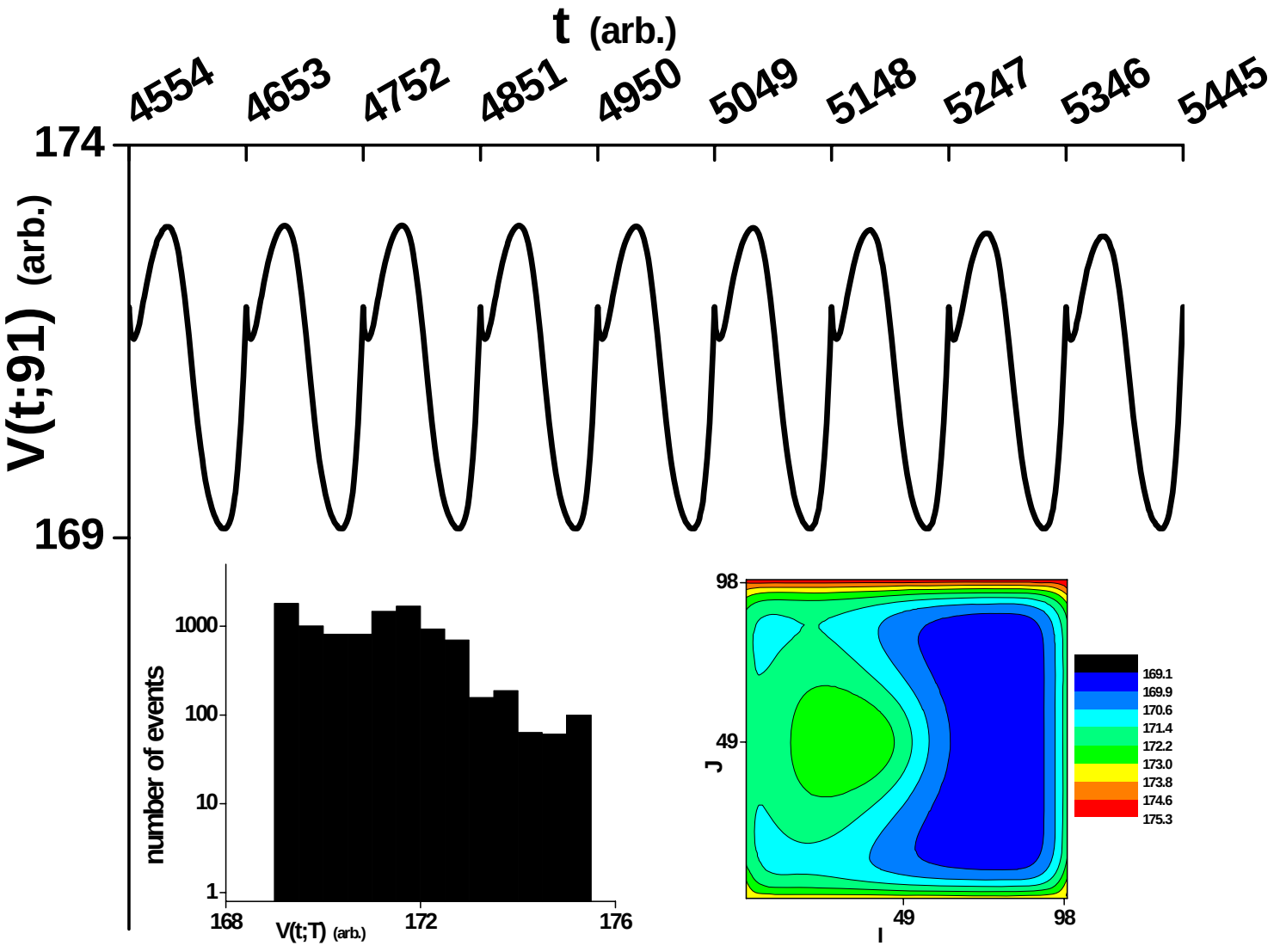

(b)

**Figure 24 (a) – (b)** Some model signals with initial 4-armed spiral waves where the amplitudes are multiplied by a Gauss factor; please see the Eqs. (19) and (20) and the related text. Please note that the time domains (4455≤t≤5445) are common in the plots: (**a**) for T=0 and (**b**) T=91 where the insets show the distributions of the number of the initial amplitudes over the mentioned amplitudes (histogram, left) and the initial amplitudes over the entries (I,J) (right), respectively.

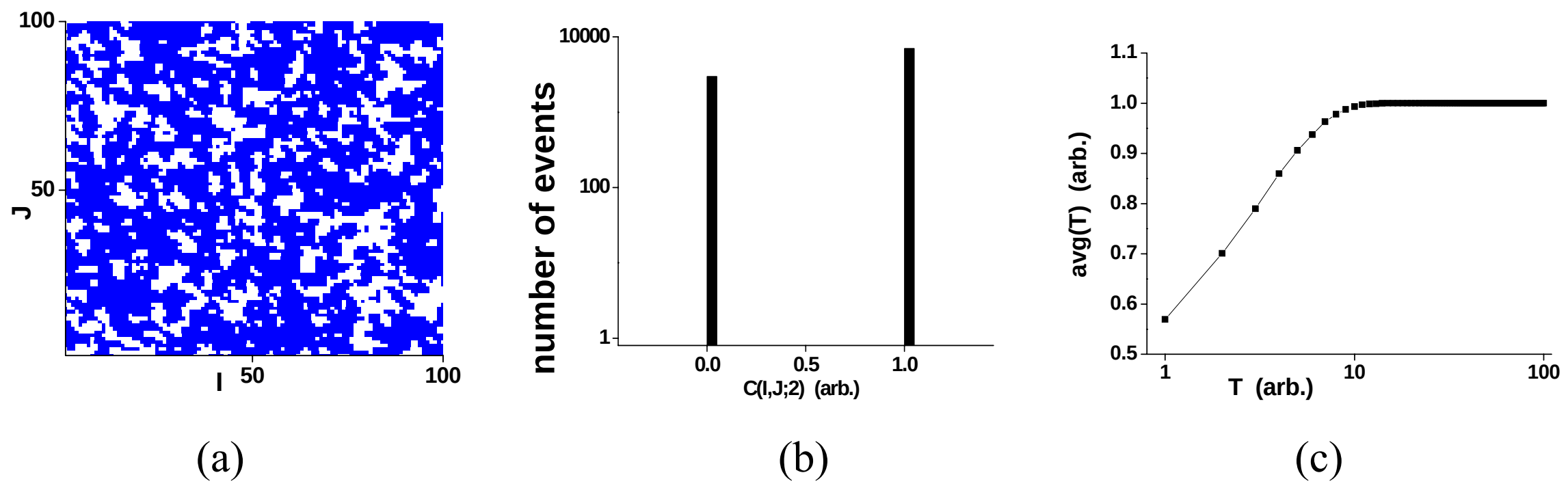

**Figure 25** The results of the present model for spatial formations for freezing (or melting) of a liquid (solid), say water, in terms of the cellular automata with the assumptions mentioned in Appendix (Sec. 5.1) and the parameters given in the Table II. (a) The distribution of the amplitudes C(I,J;T=2) over the entries of the representing lattice. The blue or white regions may be considered as the frozen portions of the surface, equally well; where, the choice for the meaning of parameter avg(T) in the Eq. (9) is decisive. avg(T) may be taken as the total supplied or extracted heat in arbitrary unit. (Please see the plot (c) here.) (b) The distribution of the number of the amplitudes (C(I,J;T=2)) over the mentioned amplitudes where the vertical axis is logarithmic. (c) The variation of the average of the (integer) amplitudes for C(I,J;T) (0<T) with time where the average (avg(T)) is same as in the Eq. (9) and the horizontal axis is logarithmic.

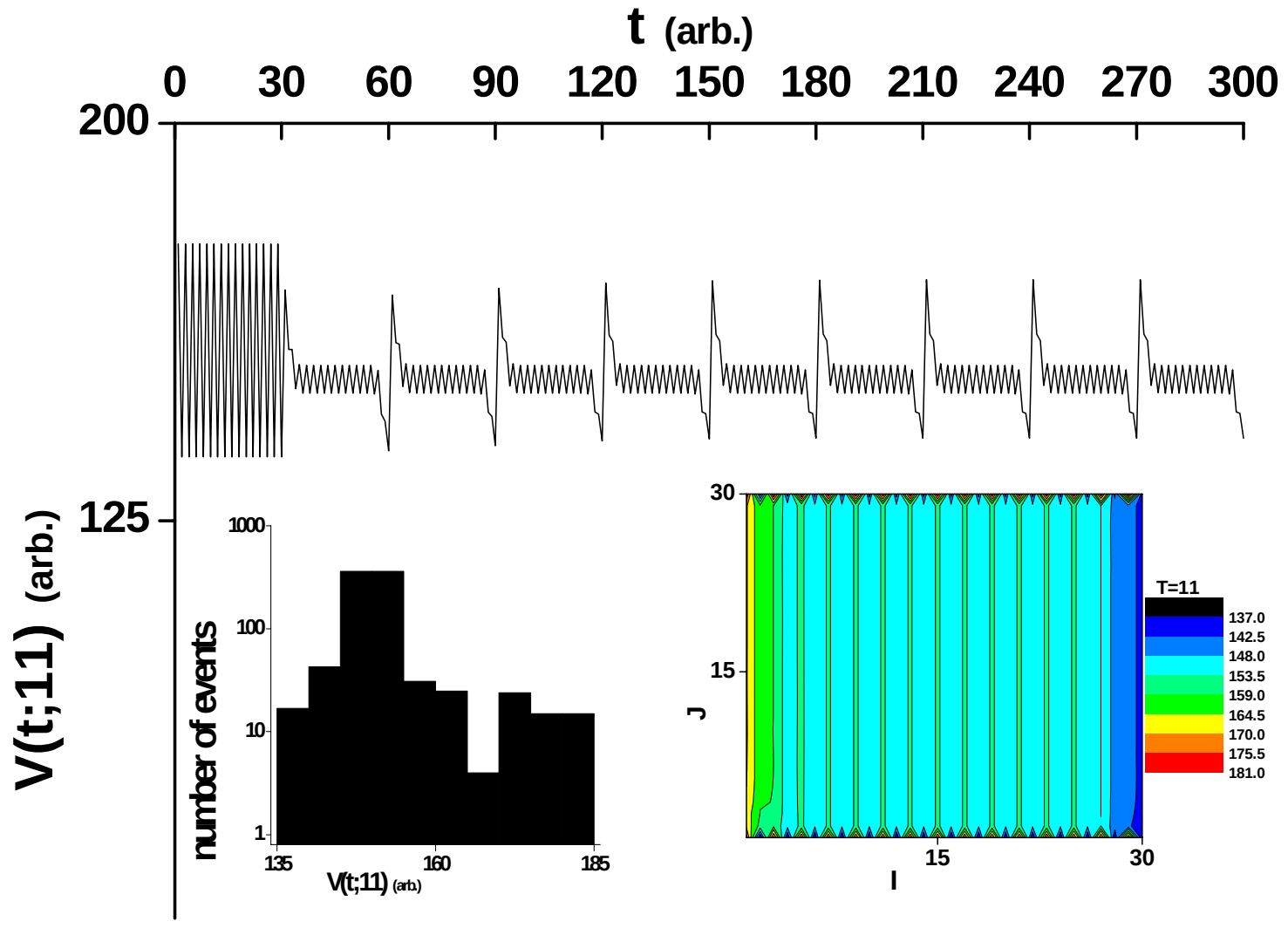

**Figure 26** The model signals with the initial wave given in the Eq. (A3) and the parameters in the Table II. The insets show the distributions of the number of the model voltage amplitudes V(t;11) over the voltages (the histogram at left, where the vertical axis is logarithmic) and the same amplitudes for C(I,J;11) over the entries (I,J) of the representing (30x30) lattice (at right), respectively. The plot is (not simulated but) computed in terms of 11 secular equations (Eq. (1)).

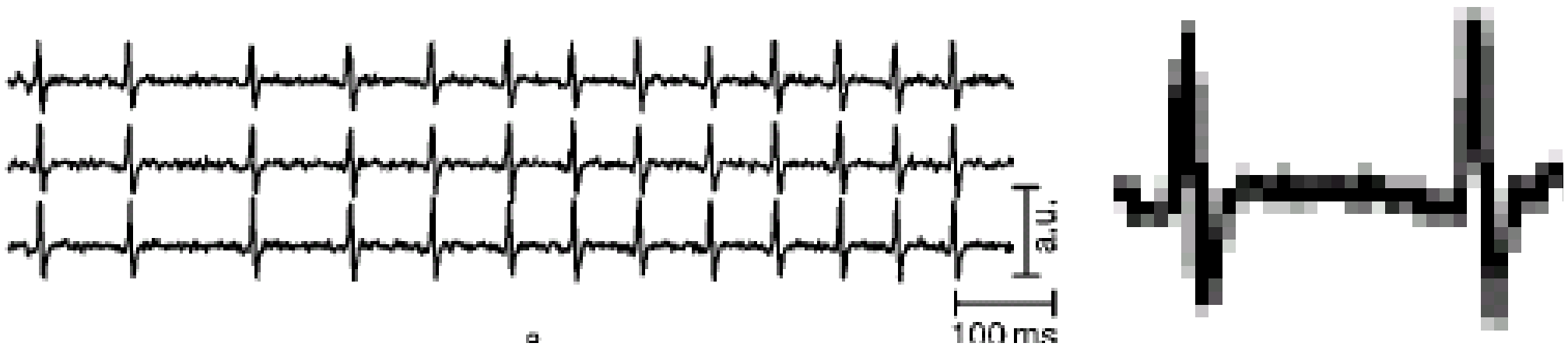

**Figure 27** The empirical EMG (left) as modified from [8] and a beat which is selected and enlarged arbitrarily.

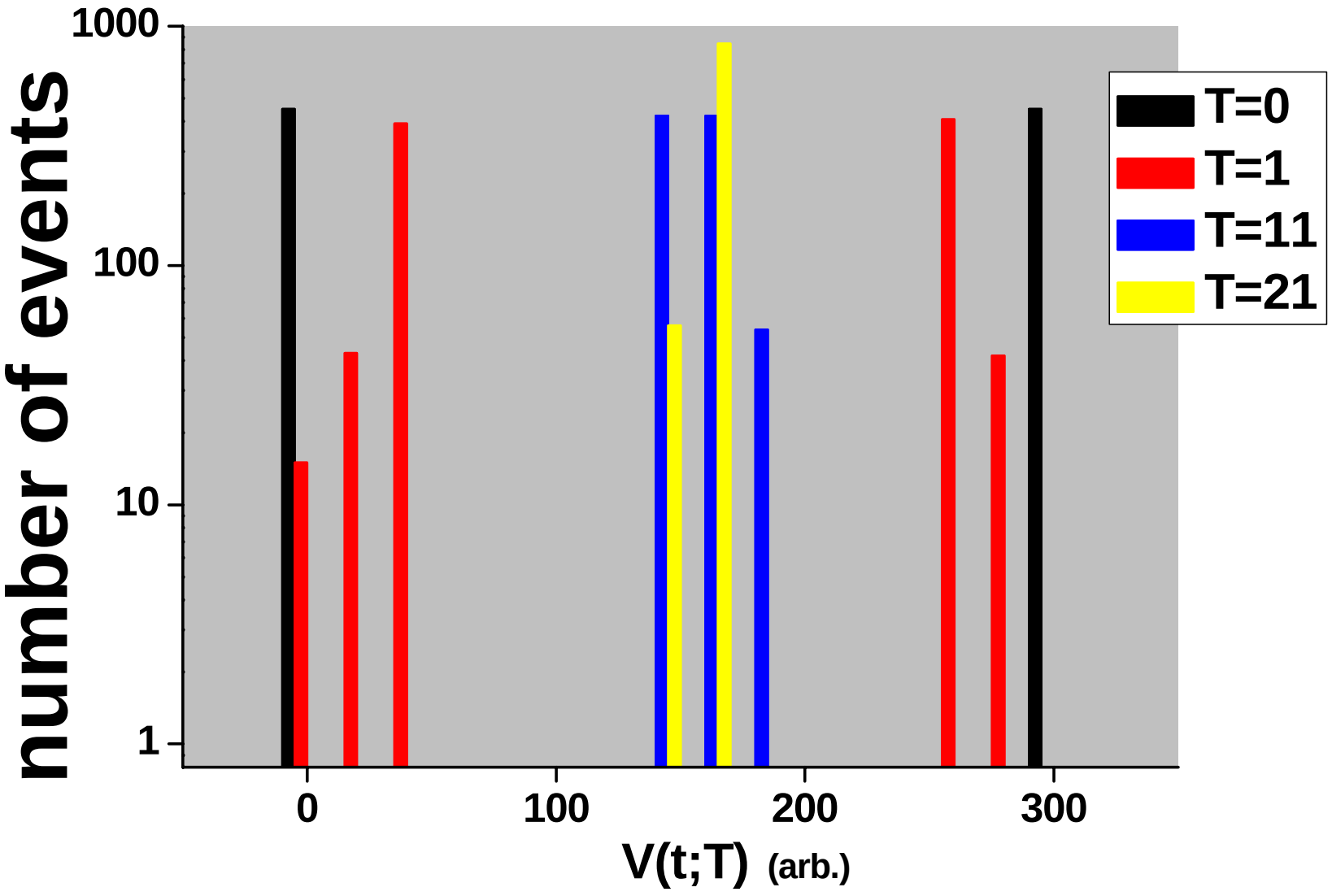

**Figure 28** Same as the left inset in The Fig. 26 but for various T and different binning.

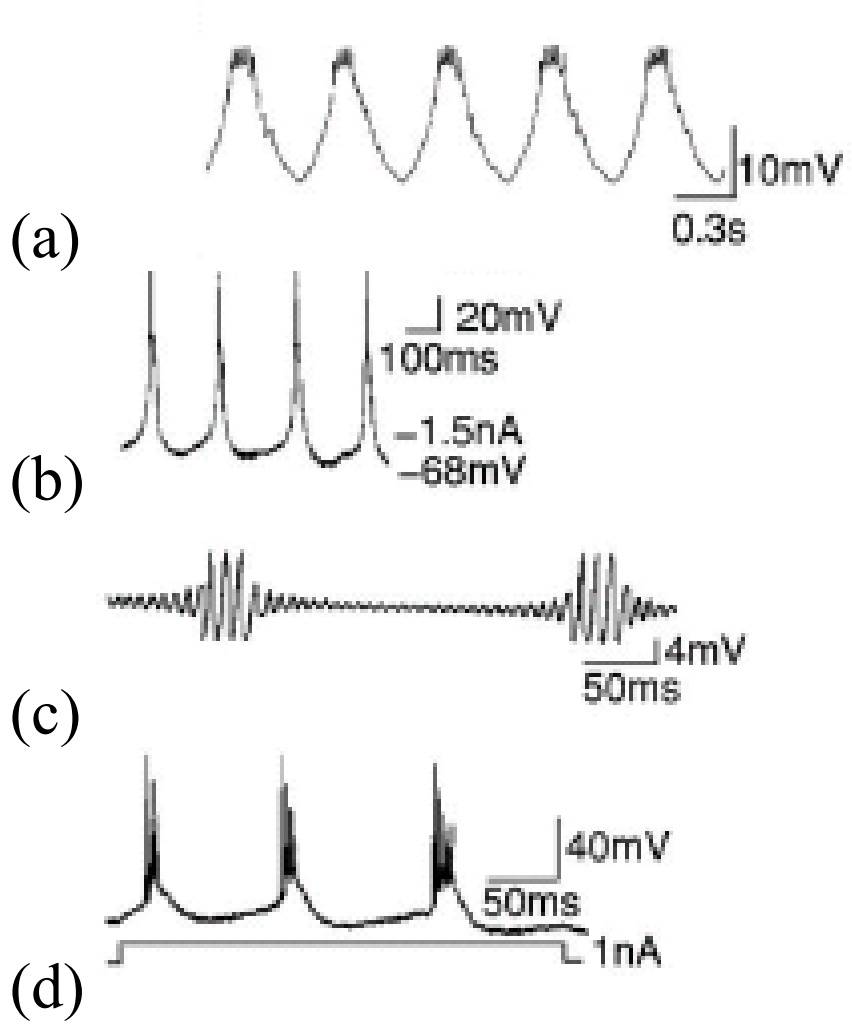

**Figure 29** The empirical data for the single neuron membrane voltage activities in (a) Lobster pyloric neuron; (b) guinea pig inferior olivary neuron; (c) sepia giant axon and (d) mouse neocortical pyramidal neuron (modified from the Figure 3 in [10] which are modified from [11]).

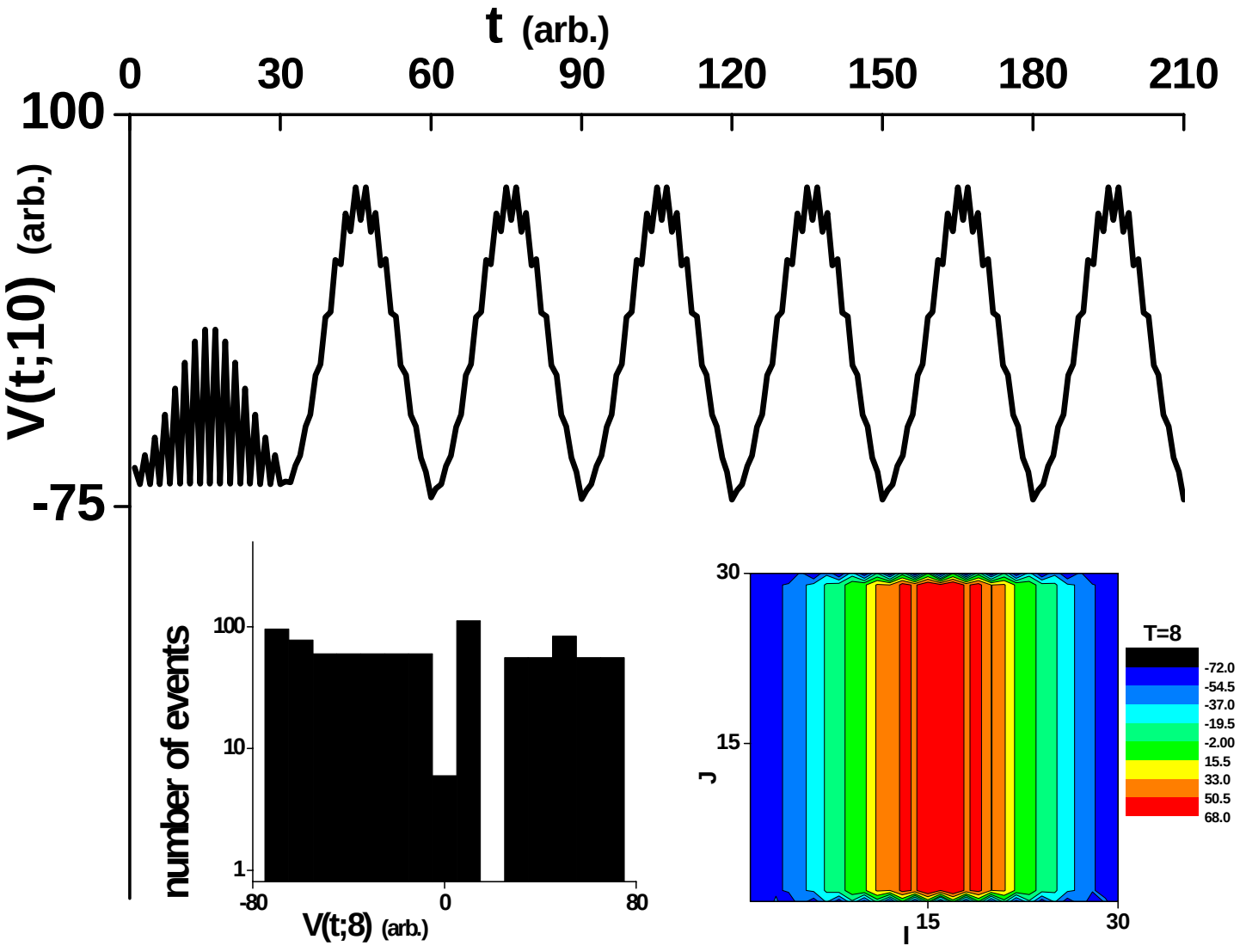

(a)

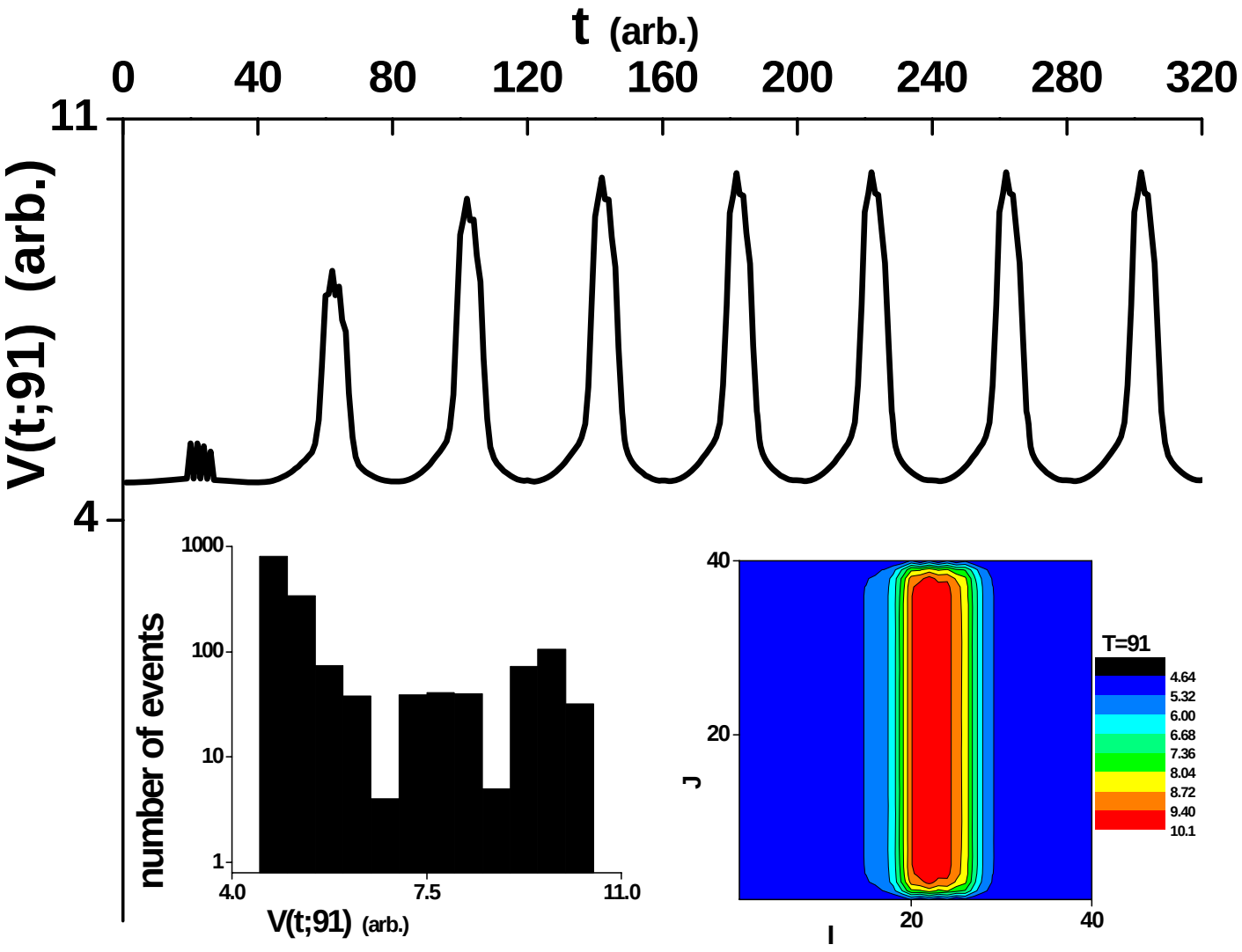

(b)

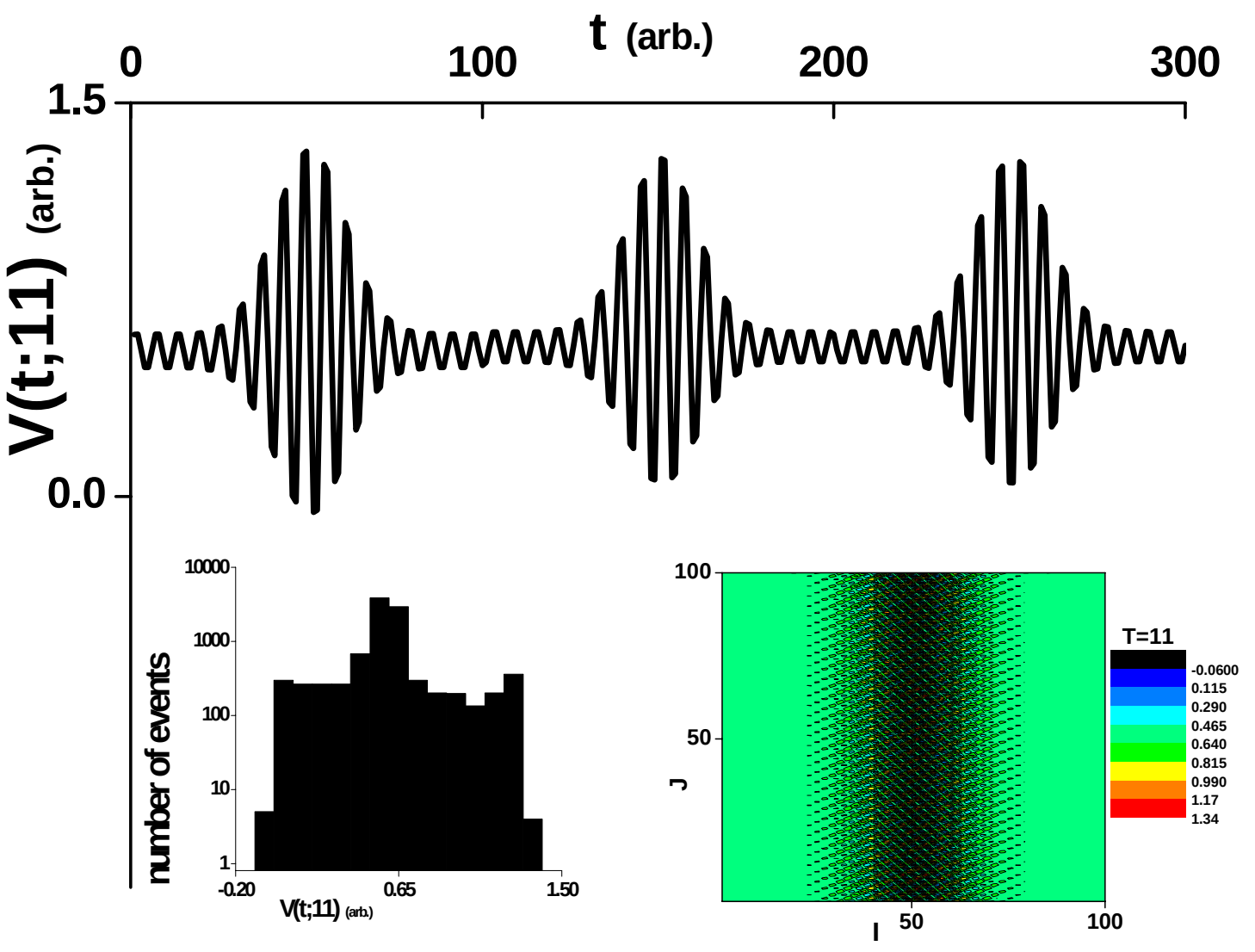

(c)

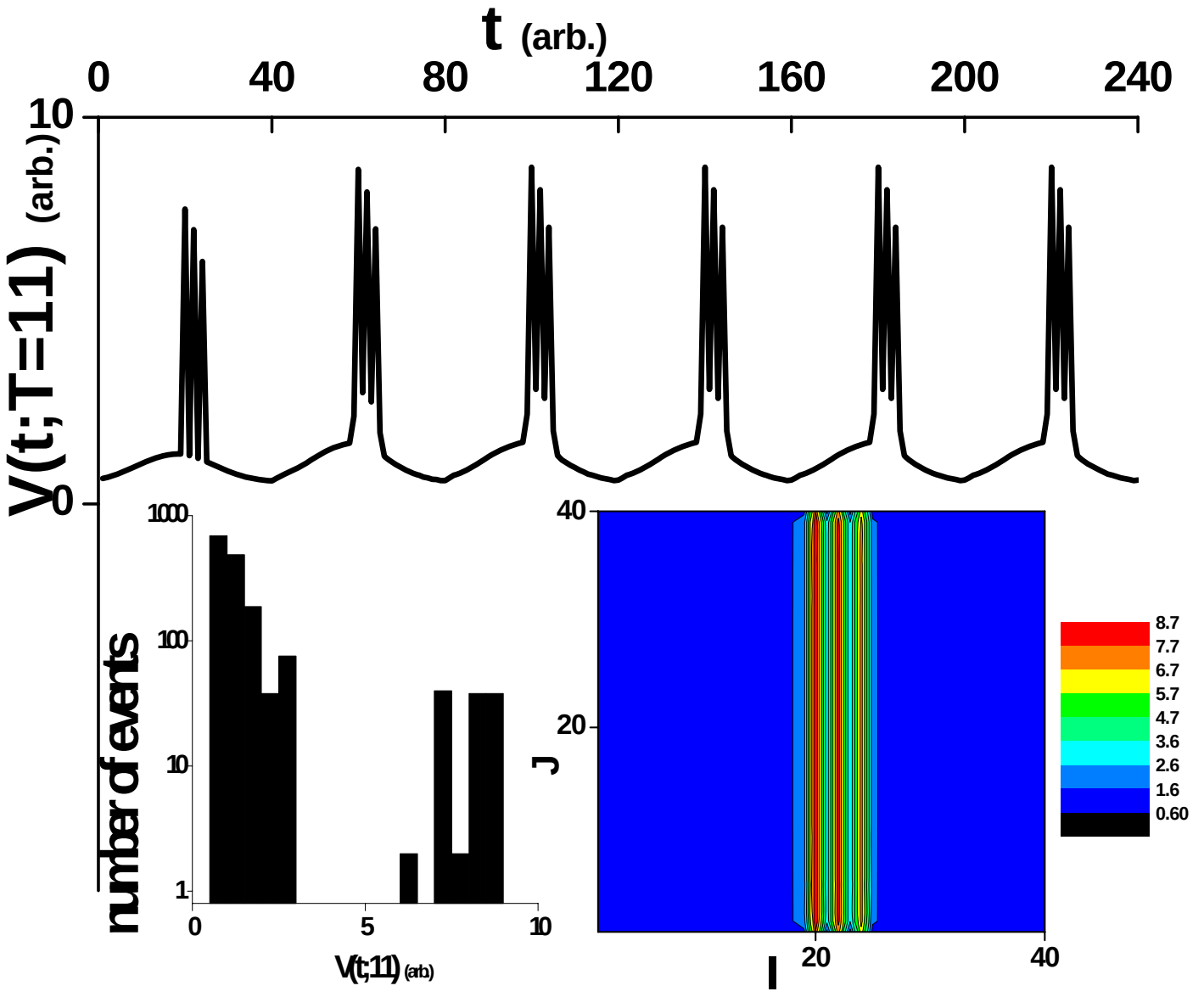

(d)

**Figure 30 (a) – (d)** Several model signals for the single neuron membrane voltage activities where the insets depict the distribution of the number of the voltages amplitudes V(t;T) over the mentioned amplitudes at the left and the same amplitudes (for C(I,J;T)) over the sites (I,J) at the right, i.e., the spatial formation, in all, respectively.